\documentclass[notitlepage,letterpaper,aps,prd,onecolumn,superscriptaddress,nofootinbib,longbibliography]{revtex4-1}
\usepackage{amsmath}
\usepackage{amsfonts}
\usepackage{amssymb}
\usepackage[latin1,utf8]{inputenc}
\usepackage{xcolor,cancel}
\usepackage[normalem]{ulem}
\newcommand{\stkout}[1]{\ifmmode\text{\sout{\ensuremath{#1}}}\else\sout{#1}\fi}
\RequirePackage{flushend}

\usepackage[colorlinks=true, pdfstartview = FitH, bookmarksopen = true, bookmarksopenlevel=2]{hyperref}
\hypersetup{pdftitle=From inflation to dark energy in scalar-tensor cosmology,pdfauthor={Jibitesh Dutta, Laur Järv, Wompherdeiki Khyllep, Sulev Tõkke}}

\usepackage[all]{hypcap}

\usepackage{latexsym}
\usepackage{graphicx}
\usepackage{color}
\usepackage{subfigure}

\begin{document}

\title{From inflation to dark energy in scalar-tensor cosmology}

\author{Jibitesh Dutta}
\email{jibitesh@nehu.ac.in}
\affiliation{Mathematics Division, Department of Basic Sciences and Social Sciences, North Eastern Hill University,  Shillong, Meghalaya 793022, India}
\affiliation{Inter University Centre for Astronomy and Astrophysics, Pune 411 007, India }

\author{Laur Järv}
\email{laur.jarv@ut.ee}
\affiliation{Laboratory of Theoretical Physics, Institute of Physics, University of Tartu, W. Ostwaldi 1, 50411 Tartu, Estonia}

\author{Wompherdeiki Khyllep}
\email{sjwomkhyllep@gmail.com}
\affiliation{Department of Mathematics, North Eastern Hill University, Shillong, Meghalaya 793022, India} 
\affiliation{Department of Mathematics, St.\ Anthony's College, Shillong, Meghalaya 793001, India}

\author{Sulev Tõkke}
\affiliation{Laboratory of Theoretical Physics, Institute of Physics, University of Tartu, W. Ostwaldi 1, 50411 Tartu, Estonia}

\begin{abstract}
The methods of dynamical systems have found wide applications in cosmology, with focus either upon inflation or upon the passage into dark energy era. In this paper,  we endeavor to capture the whole history of the universe into a dynamical system by considering generic scalar tensor gravity with radiation and dust matter fluids in flat Friedmann-Lema\^itre-Robertson-Walker spacetime. We construct the dynamical variables in such a way that the main stages of the cosmic evolution, viz.\ inflation, radiation domination, matter domination, and dark energy domination can be represented by the respective fixed points in the phase space. 
As the evolution of solutions is ruled by the sequence of these fixed points with appropriate properties, we can determine the conditions that the scalar potential and nonminimal coupling must satisfy for the model to deliver viable cosmic history in a generic manner. We illustrate the construction by a scalar field with quartic potential, with and without quadratic nonminimal coupling to curvature.

\end{abstract}

\maketitle

\section{Introduction}
The methods of dynamical systems offer a set of helpful tools to analyze and visualize the qualitative behavior of solutions of ordinary differential equations. These methods have been applied with great success also in cosmology, see Refs.\ \cite{Wainwright_book,Coley:2003mj,Bahamonde:2017ize} for general reviews and e.g.\ Refs.\ \cite{Hohmann:2017jao,Dutta:2017fjw,Zonunmawia:2018xvf,Carloni:2018yoz,Leon:2018lnd,Alho:2019jho,Humieja:2019ywy,Franco:2020lxx} for a sample of very recent papers. In a typical approach, to scrutinize the evolution for a model given before, one writes the cosmological equations as a dynamical system, finds the fixed points and from these to deduce the generic features of cosmological evolution, as the fixed points shape the course of trajectories in the phase space. In cosmology, most authors so far have focused either upon early (inflation) or late (matter to dark energy domination) stages of the cosmic history. The present work ventures beyond the usual approach in two respects.

First, we intend to describe by a dynamical system the full history of the universe, starting from the onset of inflation till the rule of dark energy. Although the idea to unify the early and late stages of the universe within a single theory is not new \cite{Nojiri:2003ft,Sotiriou:2005hu,Cognola:2007zu,Nojiri:2007as,Nojiri:2007cq,Elizalde:2010ts,Nojiri:2016vhu,Odintsov:2018nch,Odintsov:2019evb,Zhang:2019mhd,Nojiri:2019fft,Odintsov:2020nwm}, the papers that attempt to tackle the problem in the context of dynamical systems are surprisingly few \cite{Hrycyna:2010yv,Fay:2014fta,Fay:2020egt}. Yet, mapping out the global phase space should help to address some basic questions in theoretical cosmology: how common with respect to all possible initial conditions is the occurrence  of inflation (early period of accelerated expansion), how graceful is the exit from inflation into a radiation domination era, how common it is for the solutions to trace through all the eras (e.g.\ not to skip some of them, or get stuck in one of them), etc. 
In our paper, we do not claim to provide comprehensive answers to these questions, however, we offer a set of ideas how a viable cosmic history should be realized, which could serve as a basis for future detailed investigations.

Second, we slightly alter the usual mode of study which takes a model and then analyzes its phase space. Instead, we first postulate that in order to reproduce the correct history of the universe, the phase space must be endowed with fixed points corresponding to all the eras, and then investigate which models belonging to a wide family of theories are able to provide these points with appropriate characteristics. The aim is to determine the conditions on the model functions and parameters that allow to realize the full cosmic history in a generic manner for many solutions (initial conditions) within the model. It turns out the requirement most difficult to satisfy is to have the early universe dynamics ruled by a saddle point with de Sitter like expansion and with a repulsive eigendirection pointing towards the fixed point exhibiting radiation domination. The fixed points of radiation and dust matter domination are quite naturally saddles, possibly with a heteroclinic orbit between them, to guide the flow of trajectories towards the latter. Finally, there must exist an attractive de Sitter fixed point responsible for dark energy era, that collects all the available trajectories in the phase space.

Scalar-tensor gravity (STG), a family of generalized Jordan-Fierz-Brans-Dicke theories extend general relativity (GR) by introducing a scalar field nonminimally coupled to curvature, which mediates the gravitational interaction along with the usual tensor degree of freedom \cite{Will:2018bme,Fujii:2003pa,Faraoni:2004pi}. In effect, the gravitational constant becomes dynamical depending on the value of the scalar field. While astrophysical observations put quite stringent limits on the variation of the gravitational constant \cite{Uzan:2010pm,Chiba:2011bz,Ooba:2016slp}, a large class of STG models naturally accommodates it by so-called attractor mechanism, whereby the cosmological evolution from the onset of matter domination era pulls the solutions to a regime, where the scalar field relaxes around a fixed value, the effective gravitational constant stabilizes, and the theory starts to behave rather like GR \cite{Damour:1992kf, Damour:1993id, Serna:1995pi, Santiago:1998ae, Mimoso:1998dn, Serna:2002fj, Jarv:2010xm, Jarv:2010xm,Jarv:2011sm}.

Nonminimal coupling between a scalar and curvature emerges in various physical contexts, like constructions with higher dimensions or taking into account quantum corrections to a scalar field minimally coupled to gravity. STG theories have received a lot of attention in the efforts to build viable models of early universe inflation \cite{Accetta:1985du,Futamase:1987ua,Amendola:1990nn,Fakir:1990eg,Barrow:1995fj} (Higgs inflation \cite{Bezrukov:2007ep} being one of them), and late universe including dark energy \cite{Wetterich:1987fm,Perrotta:1999am,Chiba:1999wt,Boisseau:2000pr,EspositoFarese:2000ij}.  An interesting feature of these models is a possibility of superaccelerated expansion ($\dot{H}>0$, $w_{\rm eff}<-1$) without invoking phantom fields and instabilities \cite{Gunzig:2000kk,Faraoni:2001tq,Carvalho:2004ty,Perivolaropoulos:2005yv,Nesseris:2006er,Gannouji:2006jm}. See Refs.\ \cite{Clifton:2011jh,Nojiri:2010wj,Nojiri:2017ncd,Bahamonde:2017ize} for  reviews and further references. STG models can be presented in different conformal frames and parametrizations, there is a set of rules how to ``translate'' between these representations \cite{Flanagan:2004bz, Jarv:2014hma,Jarv:2015kga}. 

The dynamical system  tools have been successfully applied in the context of STG cosmology with a special focus on concrete classes of coupling functions and potentials \cite{Amendola:1990nn,Gunzig:2000ce,Gunzig:2000kk,Carvalho:2004ty,Carloni:2007eu,Jarv:2006jd,Hrycyna:2008gk,Skugoreva:2014gka}, or considering arbitrary functions \cite{Faraoni:2006sr,Jarv:2008eb,Jarv:2010zc}, and including a single matter component \cite{Charters:2001hi,Jarv:2006jd,Hrycyna:2010yv,Sami:2012uh,Hrycyna:2015eta,Hrycyna:2015vvs,Roy:2017mnz}. General geometric properties of the STG phase space were discussed in Refs.\ \cite{Abramo:2002rn,Faraoni:2005vc,Jarv:2009zf}. In \cite{Fadragas:2014mra}, a detailed analysis was performed for STG models with both radiation and matter, but focussing on late time unverse. An interesting complete history of the Universe from inflation till the dark energy era was obtained using the dynamical systems approach in Ref.\ \cite{Hrycyna:2010yv}, although here the radiation dominated epoch was generated by nonminimal coupling only. 
For an interested reader we may recommend Ref.\ \cite{Bahamonde:2017ize} as a handy introduction to the dynamical systems tools and applications in cosmology together with a comprehensive  literature review.
 
In the present work, we focus on the cosmological dynamics of the generic class of STG theories with a general non-minimal coupling function and self interacting potential. By using the methods of dynamical systems, we derive  conditions on the coupling functions and potentials in which viable cosmological evolutionary history can be achieved by a large set of solutions. In Sec.\ \ref{sec:stg_cosmology},  we introduce the key features of STG and give the cosmological equations. Then in Sec.\ \ref{sec:phase_space}, we argue for the choice of dynamical variables and build the dynamical system. The ensuing set of fixed points along with their properties is described in Sec.\ \ref{sec:fixed_points}. Equipped with this general knowledge, we analyze the examples of GR with a cosmological constant in Sec.\ \ref{sec:example_gr}, minimally coupled scalar field with a quartic potential and cosmological constant in Sec.\ \ref{sec:example_minimal}, and nonminimally coupled scalar field with the same potential in Sec.\ \ref{sec:example_nonminimal}. Finally, Sec.\ \ref{sec:discussion} presents the summary and outlook. In the appendix \ref{CMTdS1}, we apply center manifold theory for nonhyperbolic de Sitter points in minimally coupled models, which to our knowledge has not been done before in the literature.

\section{Scalar-tensor cosmology}
\label{sec:stg_cosmology}

A generic action functional for scalar field nonminimally coupled to curvature and endowed with a potential can be written as \cite{Flanagan:2004bz, Jarv:2014hma}
\begin{equation}
\label{fl_moju1}
S = \frac{1}{2\kappa^2}\int_{V_4}d^4x\sqrt{-g}\left\lbrace {\mathcal A}(\Phi)R-
{\mathcal B}(\Phi)g^{\mu\nu}\nabla_\mu\Phi \nabla_\nu\Phi - 2\ell^{-2}{\mathcal V}(\Phi)\right\rbrace 
+ S_m\left(e^{2\alpha(\Phi)}g_{\mu\nu},\chi\right) \,.
\end{equation}
Integrating over a 4-dimensional Lorentzian curved spacetime manifold $V_4$, it incorporates four arbitrary functions of the dimensionless scalar field $\Phi$: the coupling ${\mathcal A}(\Phi)$ to the curvature scalar $R$, the generic kinetic coupling of the scalar field ${\mathcal B}(\Phi)$, the self-interaction potential of the scalar field ${\mathcal V}(\Phi)$, and the conformal coupling $e^{2\alpha(\Phi)}$ between the metric $g_{\mu\nu}$ and matter fields $\chi$, while we are not considering derivative couplings or higher derivative terms.
Fixing the functions $\left\lbrace \mathcal{A},\,\mathcal{B},\,\mathcal{V},\,\alpha \right\rbrace$ gives a specific theory in certain frame and parametrization. Indeed, one of the functions can be fixed by a conformal  transformation of the metric $g_{\mu\nu} = e^{2\bar{\gamma}(\bar{\Phi})}\bar{g}_{\mu\nu}$ which sets the frame. Another function of the four can be fixed by scalar field redefinition $\Phi = \mathfrak{f}(\bar{\Phi})$ \cite{Flanagan:2004bz, Jarv:2014hma}. 
We may use a conformal transformation to choose the Jordan frame, where $\alpha \equiv 0$ and matter test particles follow geodesics. Second, we may redefine the scalar field to give the kinetic term a canonical form, $\mathcal{B}\equiv 1$. Finally, without a loss of generality, we may separate the curvature coupling function into a constant part (representing a bare gravitational constant) and a field dependent part, $\mathcal{A}=1+f(\phi)$. This leaves us with an action 
\begin{equation}
\label{fl_moju}
S = \frac{1}{2\kappa^2}\int_{V_4}d^4x\sqrt{-g}\left\lbrace (1+f(\phi))R-
g^{\mu\nu}\nabla_\mu\phi \nabla_\nu\phi - 2\ell^{-2}{V}(\phi)\right\rbrace 
+ S_m\left(g_{\mu\nu},\chi\right) \,.
\end{equation}
Here different possible choices of $f(\phi)$ and $V(\phi)$ yield different models in the scalar-tensor family, set in the Jordan frame and this particular parametrization. If the scalar field is nondynamical, the action reduces to that of GR where a constant nonzero value of the potential plays the role of the cosmological constant $\Lambda$. In the case $f(\phi)\equiv 0$ we have a scalar field minimally coupled to gravity. The theory is free of ghosts provided 
\begin{equation}
\label{eq:no_ghosts}
E(\phi) := 2(1+f(\phi)) + 3 f_{,\phi}^2 \geq 0 \,,
\end{equation}
since then in the Einstein frame, where the the tensor and scalar degrees of freedom are clearly separated, the scalar kinetic term is endowed with the correct sign \cite{EspositoFarese:2000ij, Jarv:2014hma}.

Note that we have adopted the units with $c = 1$, but do not immediately fix the values of the nonvariable part of the effective gravitational ``constant" $\kappa^2$ and keep a positive constant parameter $\ell$ which has the dimension of length, e.g.\ the Planck length.
From a ``gravity'' convention $[\kappa^2] = 1 $, it follows that $[S] = [\hbar] = L^2$ and from a ``particle physics'' convention $[S] = [\hbar] = 1 $ it follows that  $[\kappa^2] = L^2 $. In either case the scalar field $\phi$ is dimensionless.

The modified Einstein field equation obtained by varying the action \eqref{fl_moju} with respect to the metric $g_{\mu\nu}$ is given by
\begin{eqnarray}\label{efe}
 G_{\mu\nu}&=&\frac{\kappa^2}{\left(1+f(\phi)\right)}\Big[T_{\mu\nu}+\frac{1}{\kappa^2} \Big(\nabla_\mu\nabla_\nu f(\phi)-g_{\mu\nu} \Box f(\phi)+\nabla_\mu \phi \nabla_\nu \phi -\frac{1}{2} g_{\mu\nu}\nabla^\rho\phi\nabla_\rho\phi    -g_{\mu\nu} \ell^{-2} V(\phi)\Big) \Big]\,,
\end{eqnarray}
 where $ G_{\mu\nu}$ is the Einstein tensor, $T_{\mu\nu}$ is the usual matter energy momentum tensor and $\Box \equiv \nabla_{\mu}\nabla^{\mu}$. We see that in STG, the Newtonian constant is replaced by an effective gravitational ``constant''
\begin{align}
\label{eq:G_eff}
G_{\rm eff} (\phi)=\frac{\kappa^2}{\left(1+f(\phi)\right)}\, ,
\end{align}
which is scalar field dependent. Several astrophysical observations significantly constrain the space and time variations of the gravitational constant, for instance in cosmology we know that since the recombination era $\frac{|G_{\rm rec}-G_{\rm now}|}{G_{\rm now}} < 5 \times 10^{-2}$ \cite{Uzan:2010pm,Chiba:2011bz,Ooba:2016slp}. From the theoretical point of view, negative $G_{\rm eff}$ associates the scalar with a ghost and indicates an instability \cite{EspositoFarese:2000ij}. In general, diverging $G_{\rm eff}$ triggers a curvature singularity which blocks the system to evolve from an attractive gravity to antigravity regime \cite{1981SvAL....7...36S,Futamase:1987ua,Abramo:2002rn,Jarv:2008eb,Sami:2012uh}. 

On varying the action \eqref{fl_moju} with respect to the scalar field $\phi$, one obtains
\begin{align}\label{kge}
2 \Box \phi +R f_{,\phi}(\phi)-2 \ell^{-2} V_{,\phi}(\phi)=0,
\end{align}
where comma (${}_{,\phi}$) was introduced to denote a derivative with respect to the scalar field $\phi$. We may eliminate $R$ in the last equation by substituting in the trace of Eq.\ \eqref{efe}, the result is 
\begin{equation}
\label{kge_phi}
\Box \phi + \frac{2 f_{,\phi}(\phi) + 6 f_{,\phi\phi} (\phi) f_{,\phi}(\phi)}{2 E(\phi)} \nabla^\rho\phi\nabla_\rho\phi = \frac{{2}\left( (1+f(\phi)) V_{,\phi}(\phi) - 2 f_{,\phi}(\phi) V(\phi) \right)}{\ell^2 E(\phi)} + \frac{\kappa^2 f_{,\phi}(\phi)}{E(\phi)} T \,,
\end{equation}
where $T$ is the trace of the energy momentum tensor of the matter fields and $E(\phi)$ is the combination \eqref{eq:no_ghosts}. In the case of minimal coupling this is just the Klein-Gordon equation (in curved spacetime) with $V_{,\phi}$ on the r.h.s.\ and no matter contribution. In the case of nonminimal coupling two additional aspects appear. First, the scalar field dynamics is largely governed by the effective potential \cite{Skugoreva:2014gka, Jarv:2014hma,Pozdeeva:2016cja}
\begin{equation}
\label{eq:V_eff}
V_{\rm eff}(\phi) = \frac{V(\phi)}{(1+f(\phi))^2} \,,
\end{equation}
as it is $\tfrac{(1+f(\phi))^3}{E}V_{{\rm eff},\phi} $ which in Eq.\ \eqref{kge_phi} plays the same role as $V_{,\phi}$ in the minimal coupling case. (Note that the factors $1+f$ and $E$ are related to the inverse of the effective gravitational constant \eqref{eq:G_eff} and the no ghost expression \eqref{eq:no_ghosts} and should not change their sign here.)  Second, the matter fields also act as a source to the scalar field via $T$, like they perform as a source for the tensor field in the Einstein's equations \eqref{efe} via $T_{\mu\nu}$. However, conformal matter like electromagnetic radiation or ultrarelativistic particles for which the trace of the energy momentum tensor vanishes, $T=0$, do not source the scalar field even in the nonminimal coupling case. As we will see, this nuance will be significant for cosmology as well.
 
Assuming the background to be spatially flat described by the Friedmann-Lema\^itre-Robertson-Walker (FLRW) metric,
\begin{equation}
ds^2 = - dt^2 + a^2(t) d\boldsymbol{\mathrm{x}}^2 \,,
\end{equation}
and the matter content to be homogeneously and isotropically distributed radiation and dust matter fluids, 
the field equations \eqref{efe}, \eqref{kge} become \cite{Boisseau:2000pr,EspositoFarese:2000ij,Capozziello:2007iu} 
\begin{eqnarray}
\label{Fr1 eq}
3 \left(1+f(\phi) \right) H^2 &=& \kappa^2 \rho_r + \kappa^2\rho_m + \frac{\dot{\phi}^2}{2} -3H f_{,\phi}(\phi) \dot{\phi} + \ell^{-2}V(\phi) \,, \\
\label{Fr2 eq}
-2 \left(1+f(\phi) \right) \dot{H} &=& \frac{4}{3} \kappa^2 \rho_r + \kappa^2 \rho_m + \dot{\phi}^2 + f_{,\phi\phi}(\phi)\dot{\phi}^2 + f_{,\phi}(\phi)\ddot{\phi} - H f_{,\phi}(\phi) \dot{\phi}\, ,
\end{eqnarray} 
for the metric,
\begin{equation}
\label{Phi eq}
\ddot{\phi} + 3 H \dot{\phi} = 3 f_{,\phi}(\phi) \left( \dot{H} +2H^2 \right) - \ell^{-2} V_{,\phi}(\phi)\, ,
\end{equation}
for the scalar field, and
\begin{eqnarray}
\label{r continuity}
\dot{\rho}_r + 4 H \rho_r = 0 \,, \\
\label{m continuity}
\dot{\rho}_m + 3 H \rho_m = 0\, ,
\end{eqnarray}
for the radiation ($\rho_r$) and dust matter ($\rho_m$) densities, respectively.
Here, $H=\frac{\dot{a}}{a}$ is the Hubble parameter, $a$ denotes a scale factor and dot ($\,\dot{}\,$) indicates a derivative with respect of the cosmological time $t$. 

This set of equations reduces to the familiar cosmology in GR with a cosmological constant  $\Lambda$ if the evolution of the scalar field halts, $\dot{\phi}\rightarrow 0$, $\phi \rightarrow \phi_*$, such that \cite{Jarv:2014laa,Jarv:2015kga} 
 
\begin{equation}
\label{eq:GR_limit_V}
\frac{ (1+f(\phi_*)) V_{,\phi}(\phi_*) - 2 f_{,\phi}(\phi_*) V(\phi_*) }{E(\phi_*)} = 0 \,,
\end{equation}
and
\begin{equation}
\label{eq:GR_limit_M}
\frac{ f_{,\phi}(\phi_*)}{E(\phi_*)} =0 \,.
\end{equation}
The conditions \eqref{eq:GR_limit_V}, \eqref{eq:GR_limit_M} guarantee that the $\phi$ dynamics is really stopped for good and is not regenerated in Eq.\ \eqref{Phi eq} by the slope of the effective potential or matter sources, respectively. When the scalar field is constant then in the Einstein's equation \eqref{efe} only the value of the potential $V(\phi_*)$ appears, playing the role of the cosmological constant.
A simple way to satisfy \eqref{eq:GR_limit_V}, \eqref{eq:GR_limit_M} is to have $V_{,\phi}$ and $f_{,\phi}$ vanish at the same value $\phi_*$. If the model functions are suitable, the GR limit may be realized by the solutions as an asymptotic state (past or future). Not every theory in the STG family is endowed with such $f(\phi)$ and $V(\phi)$ that their extrema overlap, in such cases the GR limit could be obtained in the regime where either the potential or matter can be neglected. If the conditions \eqref{eq:GR_limit_V}, \eqref{eq:GR_limit_M} are met in a particular model, it would be convenient to have
\begin{equation}
\label{eq:GR_limit_adjust}
1+f(\phi_*)=1 \,, \qquad \ell^{-2} V(\phi_*)=\Lambda\,,
\end{equation}
as well, as in this case the effective gravitational constant \eqref{eq:G_eff} coincides with the bare gravitational constant $\kappa^2$, and the value of the potential tied with the cosmological constant. We may reparametrize the scalar field by adding a constant to $\phi$ and adjust the theory constants $\kappa^2$ and $\ell$ so that at $\phi_*$ the relations \eqref{eq:GR_limit_adjust} hold. (We can do such manipulation only for one point, i.e.\ one value of $\phi$ in a given model.)

We may introduce the relative energy densities for radiation, dust matter, and the scalar field, respectively 
\begin{align}
\label{eq:Omegas}
\Omega_r=\frac{\kappa^2 \rho_r}{3 (1+f(\phi)) H^2},\,\quad \Omega_m=\frac{\kappa^2 \rho_m}{3 (1+f(\phi)) H^2},\, \quad \Omega_{\phi}= \frac{\dot{\phi}^2}{6(1+f(\phi))H^2} - \frac{f_{,\phi}\dot{\phi}}{(1+f(\phi))H} + \frac{\ell^{-2} V(\phi)}{3(1+f(\phi))H^2}\,.
\end{align}
In the case of minimal coupling, $f(\phi)=0$, the quantities \eqref{eq:Omegas} reduce to the familiar relative densities of radiation, dust matter, and the scalar field, while in the GR limit \eqref{eq:GR_limit_V} the latter becomes the relative density of cosmological constant. 
However, in the nonminimal case, the physical interpretation of these quantities should be exercised with caution since with the involvement of nonminimal coupling their values are not necessarily and always bounded to be less and equal to one. Still, assuming radiation and dust matter to be normal with nonnegative energy densities $\rho_r \geq0$, $\rho_m\geq0$, and ruling out the effective gravitational ``constant'' \eqref{eq:G_eff} being negative, $\Omega_r$ and $\Omega_m$ are positive definite.  
Tentatively, the first two terms in the definition of $\Omega_\phi$ in Eq.\ \eqref{eq:Omegas} may be called the relative kinetic energy density of the scalar field, while the last term there allows the interpretation of relative potential energy density of the scalar field. Under the condition that the potential can not be negative, the relative potential energy density is always nonnegative, but the relative kinetic energy density can under certain conditions dip below zero in the nonminimal coupling case.

With the definitions \eqref{eq:Omegas}, the Friedmann equation \eqref{Fr2 eq} obtains a very simple form,
\begin{equation}
\label{eq:Fr1_Omegas}
1 = \Omega_r + \Omega_m + \Omega_{\phi} .
\end{equation}
From here we can define radiation domination as the situation where $\Omega_r$ is practically one while $\Omega_m$ and $\Omega_\phi$ are negligible. Similarly, the dust matter and scalar field domination mean that $\Omega_m$ and $\Omega_\phi$ dominate over the other energy densities  in \eqref{eq:Fr1_Omegas}, respectively. We can also speak of scalar kinetic domination and scalar potential domination when the first two terms in $\Omega_\phi$ or the last term in $\Omega_\phi$ overpowers the others. As mentioned before, the relative kinetic energy of the scalar field can become negative in certain situations whereby the value of the other terms can exceed one.

Additionally, we define the effective energy density and pressure of all components combined as 
\begin{eqnarray}
\rho_{\rm eff}&=& \frac{\rho_r}{1+f(\phi)} +\frac{\rho_m}{1+f(\phi)} + \frac{1}{\kappa^2 (1+f(\phi))} \left[\frac{\dot{\phi}^2}{2} -3H f_{,\phi}(\phi) \dot{\phi} + \ell^{-2}V(\phi) \right]\,,\\
p_{\rm eff}&=&\frac{1}{3}\frac{\rho_r}{1+f(\phi)}+\frac{1}{\kappa^2 (1+f(\phi))} \left[\frac{\dot{\phi}^2}{2}+2 H f_{,\phi}(\phi) \dot{\phi} + f_{,\phi\phi}(\phi)\dot{\phi}^2 + f_{,\phi}(\phi)\ddot{\phi}- \ell^{-2}V(\phi) \right]\,.
\end{eqnarray}
Then the effective equation of state pertaining to the overall universe is given by 
\begin{align}
\label{eq:w_eff}
w_{\rm eff}=\frac{p_{\rm eff}}{\rho_{\rm eff}}=\frac{\Omega_r}{3}+\Big(\frac{\dot{\phi}^2}{2}+2 H f_{,\phi}(\phi) \dot{\phi} + f_{,\phi\phi}(\phi)\dot{\phi}^2 + f_{,\phi}(\phi)\ddot{\phi}- \ell^{-2}V(\phi)\Big) \Big/ 3(1+f)H^2.
\end{align}
As expected, in the GR limit, under radiation domination $w_{\rm eff}=\tfrac{1}{3}$, under matter domination $w_{\rm eff}=0$, and under scalar potential (cosmological constant) domination $w_{\rm eff}=-1$.
For accelerated expansion of the universe ($\ddot{a}>0$)  one requires $w_{\rm eff}< -\frac{1}{3}$, while superacceleration ($\dot{H}>0$) is engendered by $w_{\rm eff}<-1$. De Sitter solution is characterized by $w_{\rm eff}=-1$ whereby the Hubble parameter remains constant and proportional to the cosmological term, $H=\sqrt{\tfrac{\Lambda}{3}}$. A minimally coupled scalar is not capable of triggering superacceleration  and is bound to show expansions with $-1 \leq w_{\rm eff} \leq 1$ only (unless it is a ghost by nature). However, nonminimal coupling allows expansions sporting $w_{\rm eff}<-1$ and without necessarily running into instability \cite{Gunzig:2000kk,Faraoni:2001tq,Carvalho:2004ty,Perivolaropoulos:2005yv,Nesseris:2006er,Gannouji:2006jm}. For example, a wide class of STG models is known to approach the GR limit in the potential domination regime wherby $w_{\rm eff}$ exhibits damped oscillations around the $-1$ value \cite{Jarv:2010xm,Jarv:2014hma}.

\section{Constructing the phase space}
\label{sec:phase_space}

In the previous section, we have presented the STG cosmological equations for flat FLRW spacetime with radiation and dust matter. Due to their nonlinear nature, it is a daunting task to try solving them analytically in full generality. However, the methods of dynamical systems offer several tools to extract qualitative information about the behavior of solutions in the system. Let us begin here by writing the equations in the form of a dynamical system, and proceed to analyze the fixed points in the next section. 

First, we write the Friedmann equation in a dimensionless form as
\begin{equation}
\label{Fr1 dimensionless}
1=\frac{\kappa^2 \rho_r}{3(1+f(\phi))H^2} + \frac{\kappa^2 \rho_m}{3(1+f(\phi))H^2}
+ \frac{{\phi'}^2}{6(1+f(\phi))} - \frac{f_{,\phi}{\phi}'}{(1+f(\phi))} + \frac{\ell^{-2} V(\phi)}{3(1+f(\phi))H^2} \,,
\end{equation}
where 
\begin{equation}
' = \frac{d }{d \ln a} = \frac{1}{H} \frac{d }{d t} = \frac{d }{d N},
\end{equation}
denotes a derivative with respect to e-folds $N=\ln a$.
Then, let us introduce the following dimensionless variables in order to convert the cosmological equations into an autonomous system of equations,
\begin{equation}
\label{variables}
x=\frac{\kappa^2 \rho_r}{3(1+f)H^2} \,, \qquad
y=\frac{\kappa^2 \rho_m}{3(1+f)H^2} \,, \qquad
z=\frac{\dot{\phi}}{H} \,, \qquad
\phi \,.
\end{equation}
These variables are constrained by the Friedmann equation \eqref{Fr1 dimensionless} which appears as
\begin{equation}\label{constraint_eqn}
1=x+y+\frac{z^2}{6(1+f(\phi))} -\frac{f_{,\phi}(\phi) z}{1+f(\phi)} + \frac{\ell^{-2} V(\phi)}{3(1+f(\phi))H^2}.
\end{equation}

This choice of variables warrants a few comments of motivation. The variables $x$ and $y$ are the usual expansion normalized variables \cite{Wainwright_book, Coley:2003mj, Bahamonde:2017ize} adopted for the setting of nonminimal coupling. They are clearly well suited for our purposes. For instance, in the radiation domination situation (when the dust matter and scalar field densities can be neglected), the Friedmann equation reduces to $1=x$. In other words, in the radiation domination setting $x$ remains constant, which is exactly what should happen at a fixed point. Note that this does not imply that the radiation density $\rho_r$ itself is constant in cosmological time $t$. In this regime both $\rho_r$ and $H$ evolve, but in a manner where $x$ is constant. Similarly, in the dust matter domination case the Friedmann constraint imposes constant $y=1$, characteristic of the dust matter fixed point. Obviously, we would not like to consider the Hubble parameter as a dynamical variable, since such choice would only show Minkowski and de Sitter solutions (constant $H$) as fixed points.

Next, one may analogously try to introduce the remaining dynamical variables in such a way that they describe the relative kinetic and potential energy densities of the scalar field \cite{Copeland:1997et,Bahamonde:2017ize}, whereas the Hubble parameter gets completely absorbed by the variables and does not occur independently in the dynamical system. While doing that we would naturally get the scalar kinetic dominated (stiff fluid) and scalar potential dominated eras as fixed points. With some luck in picking a good potential, we may also witness scaling solutions as fixed points whereby the relative energy densities of one matter component and of the scalar field scale identically as the universe expands \cite{Uzan:1999ch,Amendola:1999qq}. Such fixed point means a situation where the shares of $\Omega_{\rm r, m}$ and $\Omega_\phi$ in the Friedmann equation stay put, while $\rho_{\rm r, m}$ and $\phi$ both evolve in cosmological time. However, the drawback with the approach described above is that in order to close the equations of the dynamical system (to make the system autonomous), one has to eliminate $\phi$ by expressing it in terms of the last dynamical variable (proportional to $V$), which is not always easy or even possible \cite{Bahamonde:2017ize}. 

In this paper, we have made $\phi$ dimensionless from the beginning and opt to keep it as an independent variable. This allows us to consider any function of $f(\phi)$ or $V(\phi)$ without worrying on how to close the dynamical system. The drawback is that we will not be able to capture some interesting behaviors in the system where the scalar field evolves in cosmological time (like the scaling solutions) as fixed points. This does not mean that such behavior is not present in the system any more, it just does not manifest itself as a fixed point. Also, the physical interpretation is now a bit less straightforward, as in general, the scalar field $\phi$  itself does not carry any direct physical meaning in gravity or cosmology, only certain invariant combinations of it like the energy density or gravitational constant \eqref{eq:G_eff} are observable \cite{Jarv:2014hma}. (However, when the scalar happens to couple to other fields nongravitationally like Higgs, then its value would be relevant in particle physics, of course.)

Finally, since the scalar field enters with second derivatives in the system \eqref{Fr1 eq}-\eqref{m continuity}, we need one more variable to account for its derivative. The simplest choice might seem to treat $\dot{\phi}$ itself as a dynamical variable of the system \cite{Skugoreva:2014gka}. However, with hindsight $z=\tfrac{\dot{\phi}}{H}$ seems a better choice, since in this case the phase space will show not only regular situations with $\dot{\phi}=0$ as fixed points, but also singular situations where $\dot{\phi}$ and $H$ diverge but $z$ still vanishes can emerge as fixed points (at specific values of $\phi$). The latter turns out to be crucial in our program to associate some unstable fixed points in the phase space with inflation. In addition, we have $\phi'=z$ which makes writing out the dynamical system much more straightforward and simple. 

These four variables encode the dynamics of the cosmological equations \eqref{Fr1 eq}-\eqref{m continuity}. From the definitions \eqref{variables}, we compute the derivatives with respect to e-folds, and then substitute in $\dot{H}$ and $\ddot{\phi}$ from the field equations \eqref{Fr2 eq} and \eqref{Phi eq}, $\dot{\rho}_r$ and $\dot{\rho_m}$ from the continuity equations \eqref{r continuity} and \eqref{m continuity}, and finally $H$ from the Friedmann equation \eqref{constraint_eqn}. In the latter, we have picked the positive branch of the square root for expanding universe, and assumed $H>0$ in all situations of interest (no ``bounce'', ``turnaround'' or asymptotic Minkowski scenarios for the universe), implying we can divide by $H$ without extra concerns. As a result, we get the following autonomous system 
\begin{eqnarray}
x' &=& \frac{1}{E} \Big( (2 C_1 + B_2) z^2 x + (8 A_1+6B_{12})x^2 + (6 A_1+6B_{12})xy 
 - (10A_2 +3\frac{A_2^3}{A_1} +6A_2 B_{2})xz - (8A_1+6B_{12})x \Big)\,,
\label{dynsys x}\\
y' &=& \frac{1}{E} \Big( (2C_1 + B_2 ) z^2 y + ( 6 A_1 + 6 B_{12} ) y^2 + (8 A_1 + 6 B_{12})xy - ( 10 A_2 + 3 \frac{A_2^3}{A_1} + 6 A_2 B_2 )yz    - (6 A_1  - 3 A_2^2 \nonumber \\ && \qquad \qquad + 6B_{12} ) y \Big)\,,
\label{dynsys y}\\
z' &=& \frac{1}{E} \Big( (C_1 + \frac{1}{2} B_2)z^3 - (3 A_2 C_2 + 7A_2 - B_1 +3 A_2 B_2) z^2  + (4A_1 +3B_{12})xz + (3A_1+3B_{12})yz \nonumber \\
&& \qquad \qquad -(12 A_{12} - 6 A_1 B_1)x -(9A_{12}-6A_1 B_1) y - (6 A_1 - 9 A_2^2 + 9 B_{12}) z 
+ 12 A_{12}-6A_1 B_{1} \Big)\,,
\label{dynsys z}\\
\phi' &=& z\,,
\label{dynsys phi}
\end{eqnarray}
where the quantities
\begin{eqnarray}
A_1 = 1+f(\phi) \,, \quad & A_2 = f_{,\phi}(\phi) \,, \quad   & A_{12} = (1+f(\phi)) f_{,\phi}(\phi) \,, \nonumber \\
B_1 = (1+f(\phi))\frac{V_{,\phi}(\phi)}{V(\phi)} \,, \quad & B_2 = f_{,\phi}(\phi) \frac{V_{,\phi}(\phi)}{V(\phi)} \,, \quad & B_{12} = (1+f(\phi)) f_{,\phi}(\phi)\frac{V_{,\phi}(\phi)}{V(\phi)} \,, \nonumber \\
C_1 = 1+f_{,\phi\phi}(\phi) \,, \quad & C_2 = f_{,\phi\phi} \,, \quad &
\end{eqnarray}
were introduced for the sake of more compact notation and the overall factor $E$ is just the no ghost expression \eqref{eq:no_ghosts}.

From the outset, we note that the system \eqref{dynsys x}-\eqref{dynsys phi} has two invariant submanifolds $x=0$ and $y=0$, as vanishing $x$ implies $x'=0$ and vanishing $y$ implies $y'=0$. In physics terms,  this reflects the premise that if there is no radiation or dust matter present in the universe, these components will not be generated during further evolution. (We will come back to this premise in the discussion Sec.\ \ref{sec:discussion}.) It follows, that we can immediately deduce that global attractors of the system must lie on the intersection of the $x=0$, $y=0$ surfaces, i.e.\ in the regime where the radiation and matter contributions vanish.

The $z=0$ surface does not appear as an invariant manifold, since nonzero $y$ as well as nonzero $V_{{\rm eff}, \phi}$ (encoded in the last two terms $12A_{12}-6A_1 B_1$ of \eqref{dynsys z}) would give dynamics to the scalar field ($z'\neq 0$) even when $z$ vanishes. More precisely, in the minimally coupled case $z=0$ and $V_{,\phi}=0$ are sufficient to make $z'=0$. In the nonminimally coupled case $z=0$ and $V_{{\rm eff}, \phi}=0$ are not sufficient to imply $z'=0$, one must also include $y=0$ but not necessarily $x=0$. The latter is so because in Eq.\ \eqref{dynsys z} the free term $12A_{12}-6A_1 B_1$ also appears as a factor in front of the single $x$, hence when the former vanishes the whole $x$-term vanishes. Obviously, all these details just reflect the physics encoded in the scalar field equation \eqref{kge_phi}. As already discussed there, the nonminimally coupled scalar field is sourced by $V_{{\rm eff}, \phi}$ as well as different types of matter, except for radiation.

The various cosmological parameters can be written in terms of variables \eqref{variables} as follows 
\begin{eqnarray}
\Omega_r &=&x, \qquad \Omega_m = y, \qquad \Omega_{\phi} \equiv \frac{z^2}{3A_1} -\frac{A_2 z}{A_1} + \frac{V(\phi)}{3 A_1 H^2}=1-x-y\,,\\
 w_{\rm eff}&=&-1+2 \frac{A_1}{E} \Big((\frac{2 A_2^2}{A_1}-B_2)+(\frac{4}{3}+B_2)x+(1+B_2)y  -(\frac{4A_2}{3 A_1}+\frac{B_2\,A_2}{A_1})z+(\frac{C_1}{3A_1}+\frac{B_2}{6A_1})z^2\Big)\,.
 \label{eq:w_eff_dynsys}
\end{eqnarray}
The discussion after Eq.\ \eqref{eq:Fr1_Omegas} implies that radiation domination means $x=1, y=0$, matter domination means $x=0, y=1$, while the scalar field domination happens when $x=0, y=0$. Several limits of the effective barotropic index $w_{\rm eff}$ were observed after Eq.\ \eqref{eq:w_eff} and apply here as well.

The Friedmann constraint \eqref{constraint_eqn} puts restrictions on the domain of the phase space variables $x,y,z,\phi$. We have assumed that matter components are usual with positive definite energy densities, $\rho_r, \rho_m\geq 0$. We also expect the scalar potential to be positive definite, $V> 0$, ruling out zeros of the potential, since the current accelerated expansion suggests that at least some tiny nonzero value of the potential must be there, whichever is its actual physical origin. In addition, we only consider the positive definite values of the effective gravitational constant \eqref{eq:G_eff} with a possibility that $G_{\rm eff}$ can diverge at the boundary of the phase space. Hence, the four dimensional phase space of the system \eqref{dynsys x}-\eqref{dynsys phi} is given by 
\begin{align}\label{phase}
\Psi=\left\lbrace (x,y,z,\phi) \in \mathbb{R}^4 \, \Big| \, x,y \geq 0 ;\,\,  1-x-y-\frac{z^2}{6(1+f(\phi))} +\frac{f_{,\phi}(\phi) z}{1+f(\phi)} > 0 \right\rbrace.
\end{align} 

As the phase space \eqref{phase} is unbounded in some of the directions, we may lose sight of the interesting asymptotic features of the system pertaining to the infinite values of the variables. In order to have a complete picture on the global dynamics, we can compactify the phase space by applying  the following Poincar\'e's transformation:
\begin{align}
\label{poinc}
x_p = \frac{x}{r} \,, \quad
y_p = \frac{y}{r} \,, \quad
z_p = \frac{z}{r} \,, \quad
\phi_p=\frac{\phi}{r}\,, \qquad
r=\sqrt{1+x^2+y^2+z^2+\phi^2} \,.
\end{align}
In the compactified phase space
\begin{eqnarray} 
\Psi_p &=&\left\lbrace (x_p,y_p,z_p,\phi_p) \in \mathbb{R}^4 \, \Big| \, 0 \leq x_p,y_p \leq 1; \,\, -1\leq z_p,\phi_p \leq 1, 0\leq x_p^2 + y_p^2+ z_p^2+\phi_p^2 \leq 1\,, \right. \nonumber \\&& \left.  1 - \frac{x_p}{r_p}-\frac{y_p}{r_p}-\frac{z_p^2}{6 r_p^2 \left(1+f(\phi)|_{\phi=\frac{\phi_p}{r_p}}\right)}+\frac{ z_p\, f_{,\phi}(\phi)|_{\phi=\frac{\phi_p}{r_p}}}{r_p \left(1+f(\phi)|_{\phi=\frac{\phi_p}{r_p}}\right)}>0 \right\rbrace \,,
\label{comp_phase}
\end{eqnarray} 
with $r_p=\sqrt{1-x_p^2-y_p^2-z_p^2-\phi_p^2}$. The points at infinity in $\Psi$ are associated with points on  unit hypersphere
\begin{eqnarray}\label{comp_sub_phase}
\left\lbrace (x_p,y_p,z_p,\phi_p) \in \Psi_p \, \Big| \, x_p^2 + y_p^2 +  z_p^2+\phi_p^2 = 1 \right\rbrace\,.
\end{eqnarray} 
The equations for the compactified dynamical system in $x_p, y_p, z_p, \phi_p$ can be obtained by using the transformation \eqref{poinc} on the system \eqref{dynsys x}-\eqref{dynsys phi}. We will not give the resulting equations here explicitly due to their extremely lengthy and unwieldy expressions. However, we will refer to these equations in the discussion of asymptotic fixed points, and use them numerically in plotting the trajectories for concrete examples in the following sections.

\section{Fixed points and their properties}
\label{sec:fixed_points}

In order to understand the behavior of the solutions in the system \eqref{dynsys x}-\eqref{dynsys phi}, we will carry out the dynamical systems analysis in a standard manner \cite{Wainwright_book,Coley:2003mj,Bahamonde:2017ize}. We extract the fixed points from the equations, compute the eigenvalues of the perturbed matrix around these fixed points, and find the eigenvectors corresponding to each eigenvalue. We also discuss the fixed points at infinity. In this section, we present the main mathematical results while the next three sections discuss more physics by looking at concrete examples with increasing complexity. 

The fixed points (also ``equilibrium points'') of a dynamical system correspond to the solutions with no dynamics in terms of the variables of the system, i.e.\ 
\begin{equation}
\label{eq:fixed_point_condition_general}
x'=0, \quad y'=0, \quad z'=0, \quad \phi'=0 \,.
\end{equation}
For a general coupling function $f(\phi)$ and general potential $V(\phi)$, the dynamical system \eqref{dynsys x}-\eqref{dynsys phi} has three classes of fixed points which entertain different cosmological eras: 
points corresponding to radiation domination (R) at $x=1, y=0, z=0$, points corresponding to matter domination (M) at $x=0, y=1, z=0$, and points corresponding to scalar field domination $x=0, y=0, z=0$ which behave effectively like de Sitter (dS). In certain cases, the phase space may also contain some other fixed points where the variables assume other values, but these cases depend on the specifics of the functions  $f(\phi)$ and $V(\phi)$ which makes their general assessment cumbersome and not very informative. In the following, let us briefly characterize each of the three classes of points mentioned above.
\begin{enumerate}
\item[$R$] The radiation dominated points $(1,0,0,\phi_r)$ exist for any choice of the coupling function $f(\phi)$ and potential $V(\phi)$. In fact, they form a line in the phase space since the condition \eqref{eq:fixed_point_condition_general} is satisfied in radiation domination for any value of the scalar field, $\phi_r$ (such feature does also occur when one employs some other definitions of dynamical variables \cite{Damour:1993id,Jarv:2006jd,Roy:2017mnz}). The effective barotropic index at these points, $w_{\rm eff}=\frac{1}{3}$, takes the usual value for relativistic matter.

The eigenvalues of the corresponding perturbed matrix are
\begin{equation}
4 \,, \quad 1 \,, \quad  0\,, \quad -1 \,,
\end{equation}
with the corresponding eigenvectors 
\begin{equation}
\left(\begin{array}{c}
4\\
0\\
4\,r_1\\
r_1 \end{array} \right)\,, \quad
\left(\begin{array}{c}
-1\\
r_2\\
r_3\\
r_3 \end{array} \right), \quad
\left(\begin{array}{c}
0\\
0\\
0\\
1 \end{array} \right) \,, \quad
\left(\begin{array}{c}
-\frac{f_{,\phi}}{1+f}\\
0\\
-1\\
1\end{array} \right)\,,
\end{equation}
where 
\begin{eqnarray}
r_1&=&\frac { 6 \left( 1+f \right) 
 \left(  V_{,\phi}  \left( 1+f \right) -2\, f_{,\phi} V \right)}{6\, V_{,\phi}  f_{,\phi} \left( 1+f \right) +3\, f_{,\phi}^{2} V +10\, \left( 1+f \right) V}\,,\\
r_2&=&\frac {4(1+f)+6f_{,\phi}^2}{4\, (1+f) + 3f_{,\phi}^2}\,,\\
r_3&=&\frac {3\, (1+f) f_{,\phi}}{4\, (1+f) + 3f_{,\phi}^2}\,,
\end{eqnarray}
and all the quantities must be evaluated at the value $\phi_r$ under consideration. We see that all the fixed points along the fixed line maintain the same characteristic of a saddle, where two directions are repulsive to the solutions (positive eigenvalues) and one is attractive (negative eigenvalue), although the precise orientation of the eigendirections changes point by point. The eigenvector corresponding to the zero eigenvalue is aligned along the fixed line itself, as should be the case, since it is impossible for the solutions to evolve along the fixed line. The attractive eigenvector (corresponding to the negative  eigenvalue) lies on the surface $y=0$ where dust matter vanishes. This implies that trajectories tend to approach the radiation domination point in a regime where dust matter component is negligible. The first repulsive eigendirection does also reside on the $y=0$ surface, and guides the solutions away from the radiation domination to the de Sitter attractor when dust matter is not present. In the case of minimal coupling, the second repulsive eigenvector lies on the $x-y$ plane and points directly towards the the matter domination  fixed point. In the case of nonminimal coupling  this eigenvector can deviate from the $x-y$ plane. This tells that solutions with at least some tiny amount of dust matter present will evolve away from radiation domination towards matter domination or scalar field domination, depending on the model and initial conditions.

\item[$M$] The dust matter dominated points $(0,1,0,\phi_m)$ exist when $\phi_m$ satisfies 
\begin{equation}
\label{M fixed point condition}
f_{,\phi}(\phi_m)=0 \,.
\end{equation}
It means in the minimally coupled case the dust matter dominated points form a fixed line in the $\phi$ dimension like the radiation dominated points. However, in the case of nonminimal coupling, the dust matter dominated points exist only at certain values of $\phi$, or may not exist at all in a given model, if the coupling $f(\phi)$ has no extrema. This is a reflection of the ``attractor mechanism'' whereby in the regime where the dust matter dominates and the scalar potential can be neglected, the scalar field will rest only at the GR limit \eqref{eq:GR_limit_M} \cite{Damour:1993id, Santiago:1998ae,  Jarv:2011sm}.
From Eq.\ \eqref{eq:w_eff_dynsys}, we see that all dust matter fixed points correspond to an unaccelerated universe with $w_{\rm eff}=0$ like in GR. 

The eigenvalues of the perturbed matrix are 
\begin{equation}
 3, \quad   -\frac{3}{4} \left[1  +  \sqrt{1+\frac{8}{3} f_{,\phi\phi}} \right], \quad  -\frac{3}{4} \left[1  -  \sqrt{1+\frac{8}{3} f_{,\phi\phi}} \right] , \quad -1 \,,
\end{equation}
and the corresponding eigenvectors are given by 
\begin{equation}
\left(\begin{array}{c}
0\\
1\\
3 m_1\\
m_1 \end{array} \right)\,, \quad 
\left(\begin{array}{c}
0\\
0\\
m_{2+}\\
1 \end{array} \right) \,, \quad 
 \left(\begin{array}{c}
0\\
0\\
m_{2-}\\
1 \end{array} \right) \,, \quad
\left(\begin{array}{c}
-1\\
1\\
0\\
0 \end{array} \right)\,, 
\end{equation}
where
\begin{eqnarray}
m_1 &=&\frac{2 (1+f)V_{,\phi}}{(9-f_{,\phi\phi}) \, V(\phi)} \,, \\
m_{2\pm}&=&\frac {-3\pm\sqrt {9+24 f_{,\phi\phi} }}{4} \,,
\end{eqnarray}
and all quantities must be computed at the value ${\phi=\phi_m}$ under consideration. 

The fixed point $M$ has the character of a saddle with one or two repulsive eigendirections and three or two attractive eigendirections, depending on the model function $f(\phi)$. In the case of minimal coupling where the M fixed points form a line since $\phi_m$ is left arbitrary, the third eigenvalue vanishes and the related eigenvector points along this line, like in the case of the R points.
The fourth eigenvector that corresponds to an attractive eigenvalue lies on the $x-y$ plane and attracts the trajectories coming from the radiation domination point. In the case of minimal coupling, the repulsive eigendirection of $R$ and attractive eigendirection of $M$ align, indicating a heteroclinic orbit between them (a trajectory running from one fixed point to another).
 However, the eigenvector that corresponds to another attractive eigenvalue does not lie on the $x-y$ plane which implies that also trajectories coming from scalar field domination or infinity approach matter domination without passing through radiation domination. Further, the eigenvector that corresponds to the first repulsive eigenvalue lies on the surface $x=0$. This signifies that the trajectories which repel away from this point correspond to vanishing radiation solutions and hence must be attracted towards a de Sitter point.
 
In a system with general coupling function, but without radiation and potential, approximate analytic solutions near this fixed point can be found in Ref. \cite{Jarv:2011sm}.

\item[$dS$] The scalar field dominated points $(0,0,0,\phi_*)$ exist at the values $\phi_*$ where 
\begin{equation}
\label{dS fixed point condition}
\frac{6 (1+f(\phi_*))}{E(\phi_*)} \left( 2f_{,\phi}(\phi_*) -(1+f(\phi_*)) \frac{V_{,\phi}(\phi_*)}{V(\phi_*)} \right)=0 \,,
\end{equation}
and recall we have assumed $V(\phi)\neq 0$. Again the existence of such points in a given model depends on the functions $f(\phi)$ and $V(\phi)$. For the regular values of the potential, the condition \eqref{dS fixed point condition} is satisfied at the extrema of $V_{\rm eff}$ \eqref{eq:V_eff}, in the case of minimal coupling at the extrema of $V$. Moreover, by comparing with Eq.\ \eqref{eq:GR_limit_V} we see the existence of this fixed point coincides with the GR limit. In fact, the latter is not an arbitrary coincidence, but derives from how we have set up our variables and constructed our system. In the discussions of Sec. \ref{sec:stg_cosmology}, we defined the GR limit occurring at the value where the scalar field feels no force, which in the absence of matter is provided by the extremum of the effective potential. On the other hand, the absence of force is just the definition of a fixed point. As far as expansion is concerned, this point corresponds to an accelerated universe with de Sitter behavior $w_{\rm eff}=-1$. 

The eigenvalues of the perturbed matrix are 
\begin{align}
\label{eq:dS_general_eigenvalues}
 -\frac{3}{2} \pm \frac{3}{2}\sqrt{1+\frac{8}{3E} \left( 2 (1+f) f_{,\phi\phi} +2 f_{,\phi}^2 - (1+f)^2 \frac{V_{,\phi\phi}}{V} \right) },\, -3,\, -4,
\end{align}
associated with the eigenvectors (taking first `$+$' and then `$-$' in the above)
\begin{align}
\label{eq:dS_general_eigenvectors}
\left(\begin{array}{c}
0\\
0\\
d_{1+}\\
1 \end{array} \right) \,, \quad
\left(\begin{array}{c}
0\\
0\\
d_{1-}\\
1 \end{array} \right) \,, \quad
 \left(\begin{array}{c}
0\\
d_2\\
-3 \, d_3\\
d_3 \end{array} \right)\,, \quad
\left(\begin{array}{c}
1\\
0\\
0\\
0 \end{array} \right)\,,
\end{align}
where 
\begin{eqnarray}
 d_{1\pm}&=&\frac {24 \left( 1+f  \right) f_{,\phi\phi} V  +24 f_{,\phi}^{2}V  -12 \left( 1+f  \right) ^{2}\,V_{,\phi\phi} }{3 E V \pm \sqrt {3}\sqrt {A}}\,, \label{d1}\\
 \frac{d_3}{d_2}  &=&  \frac{ \left( 1+f  \right) f_{,\phi} V }{ -4\, f_{,\phi\phi} V   \left( 1+f  \right) -4\, f_{,\phi} ^{2}V +2\, \left( 1+f  \right) ^{2}\,V_{,\phi\phi}} \,, \label{d2} \label{d3} \\
A&=& E V \Big( 16 \left( 1+f \right) f_{,\phi\phi} V   +25 f_{,\phi}^{2}V + \left( 6+6 f   \right) V -8 \left( 1+f  \right) ^{2}V_{,\phi\phi} \Big)\,,
\end{eqnarray}
and
$E$ was defined at the no ghost condition \eqref{eq:no_ghosts} and all the quantities must be evaluated at the value $\phi_*$ under consideration.

In the case of minimal coupling  the situation is quite clear. The dS point is an attractor (all eigenvalues negative) when it  occurs at a local minimum of the potential, $V_{,\phi\phi}({\phi_*})>0$, and a saddle with one repulsive eigendirection when it occurs at a local maximum of the potential, $V_{,\phi\phi}(\phi_*)<0$. In particular, it is a stable node for $0 < \tfrac{V_{,\phi \phi}}{V} < \tfrac{4}{3}$ and a stable focus (spiral) when $\tfrac{V_{,\phi \phi}}{V} > \tfrac{4}{3}$.  The linear stability analysis is inconclusive if $V_{,\phi\phi}(\phi_*)=0$, but in such case, the method of center manifold theory can be employed to determine the behavior near the corresponding center manifold of the point. For minimal coupling,  we present such analysis in the appendix \ref{CMTdS1}.

In the particular subcase where the minimally coupled scalar field has a constant potential (i.e.\ there is only a cosmological constant), the condition \eqref{dS fixed point condition} does not fix the value of $\phi_*$ and the dS fixed points form a line in the $\phi$ dimension of the phase space (like the $R$ points). In such a situation, the eigenvalues reduce to $0, -3, -3, -4$ and the corresponding eigenvectors are
\[
\left(\begin{array}{c}
0\\
0\\
0\\
1 \end{array} \right) \,, \quad
\left(\begin{array}{c}
0\\
0\\
-3\\
1 \end{array} \right) \,, \quad
 \left(\begin{array}{c}
0\\
1\\
0\\
0\end{array} \right)\,, \quad
\left(\begin{array}{c}
1\\
0\\
0\\
0 \end{array} \right)\,.
\]
Note that the first eigenvector corresponding to the zero eigenvalue points along the $\phi$ dimension and signifies no dynamics along the line of fixed points.

In the nonminimal coupling case the dS fixed points have the character of an attractor, unless
\begin{equation}
\label{dS fixed point eigenvalue positive}
\frac{2(1+f)f_{,\phi\phi} + 2 f_{,\phi}^2 - (1+f)^2 \tfrac{V_{,\phi\phi}}{V} }{\left(2(1+f)+3f_{,\phi}^2 \right) }  >0 \,,
\end{equation}
where one of the eigenvalues is positive. In fact, the condition \eqref{dS fixed point eigenvalue positive} can be reconstructed as
\begin{equation}
\frac{(1+f)^3}{E} \frac{V_{{\rm eff},\phi\phi}}{V} <0 \,,
\end{equation}
with the fixed point condition \eqref{dS fixed point condition} substituted in. This accords with our observation after Eq.\ \eqref{eq:V_eff} that in the nonminimal coupling case the scalar field dynamics is largely governed by the effective potential. The fixed point occurs at the local extremum of the effective potential, it is an attractor at the local minimum and a saddle at the local maximum of the effective potential. 
If $V_{{\rm eff},\phi\phi}$ vanishes, this point becomes nonhyperbolic and full analysis would again require the application of the center manifold theory, as in the minimally coupled case (in appendix \ref{CMTdS1}). However, we will not perform this analysis here, since in the general case the formulas become rather unwieldy and not very informative.

In \eqref{eq:dS_general_eigenvectors}, the first eigendirection is the only one that can be repulsive, it is orthogonal to the $x-y$ plane and guides solutions towards or from a fixed point at another value of $\phi$. The second eigendirection is attractive and also orthogonal to the $x-y$ plane. Therefore, when a model allows e.g.\ two dS fixed points, one saddle and one attractor, most likely there will be a heteroclinic orbit between them, flowing from the saddle to attractor along the plane where radiation and matter are zero. The third eigendirection is also attractive and pulls the solutions from the side of the radiation dominated point $R$. The fourth eigendirection is attractive and collects the solutions from the matter dominated point $M$. One would expect that an unstable dS saddle is very useful in engendering reasonable cosmic history, providing a natural setting for inflation. Indeed, the three attractive eigendirections would gather many solutions from the phase space into the vicinity of this point, these solutions would then undergo almost de Sitter type accelerated expansion as they pass by the fixed point, and then leave the neighborhood of this point following the first (repulsive) eigendirection, whereby the inflationary expansion terminates. For an extensive set of initial conditions, this point would provide a transient phase of accelerated expansion. However, there is a complication. The first eigendirection does not immediately point to the radiation dominated point, that according to expectations should take over the relay baton in the cosmic race. Instead, if the system has besides a saddle dS also an attractor dS, many of the solutions would flow directly from the saddle to the stable node dS, not giving much chance for radiation or matter to increase their share in the cosmic budget. Yet, both radiation and matter dominated points were endowed with attractive eigendirections, these could still pull a chunk of trajectories into their zones of influence. How this scenario precisely unfolds in the phase space depends on the relative positions of the fixed points. In the general context we can at best state that for the dS point to play a role in inflation, it has to be a saddle, i.e.\ the condition \eqref{dS fixed point eigenvalue positive} must hold. Otherwise, the dS point is a stable node, probably responsible for dark energy. As the dS attractor lies on the intersection of invariant submanifolds, it is, therefore, representing a global attractor irrespective of the choice of initial conditions.

In a system with general nonminimal coupling and potential, but without radiation and dust matter this fixed point was described before in Refs.\ \cite{Faraoni:2006sr, Jarv:2008eb, Jarv:2010zc}, and the analytic form of approximate solutions near it can be found in Refs.\ \cite{Jarv:2010xm, Jarv:2014hma}.

\item[$dS_{\infty}$] A closer consideration of the dS point condition \eqref{dS fixed point condition} reveals, that it can also exist at the scalar field infinity, $|\phi| \rightarrow \infty$. Indeed, by inspecting the system \eqref{dynsys x}-\eqref{dynsys phi} for minimal coupling, we see that the limit $x=0$, $y=0$, $z=0$, $|\phi|\rightarrow \infty$ can imply $x'=0$, $y'=0$, $z'\rightarrow 0$, $\phi'=0$, i.e.\ there is a fixed point, provided 
\begin{equation}
\frac{V_{, \phi}}{V}\Big|_{|\phi| \rightarrow \infty} \rightarrow 0 \,.
\end{equation}
For instance, this condition is satisfied for power law potentials.   
In the nonminimal coupling case, a model with quadratic coupling function $f=\xi \phi^2$ and quartic potential $V=V_0 \phi^4 + \Lambda$ is endowed by such an asymptotic fixed point, for example. Other models possessing it can be also constructed. The eigenvalues characterizing the point need to be properly recalculated in the compact variables \eqref{poinc} with considerable extra effort, however, quick numerical checks of the evolution near this point in the compact phase space \eqref{comp_phase} reveal that the point is a saddle when the effective potential \eqref{eq:V_eff} has an asymptotic maximum there, and an attractor, when it has an asymptotic minimum.

It is interesting that in this regime we have de Sitter like accelerated expansion with $w_{\rm eff}=-1$, but neither the scalar field nor the Hubble parameter are constant. In fact, the Friedmann constraint \eqref{constraint_eqn} allows 
$V\rightarrow \infty$,
$H\rightarrow \infty$ while $z=\tfrac{\dot{\phi}}{H} \rightarrow 0$.
 Hence strictly speaking we might better call these points ``asymptotic de Sitter'' or ``quasi de Sitter'', in contrast with the true de Sitter points where $H$ is a constant. Such asymptotic behavior of the solutions has been noted before \cite{Skugoreva:2014gka}, but not in connection with a fixed point. Our choice of dynamical variables brings out this regime as a fixed point.  One should also note, that contrary to the regular $dS$ fixed point, the asymptotic $dS_\infty$ point does not satisfy the condition \eqref{eq:GR_limit_V} of GR with a cosmological constant. Nevertheless, it is very useful in the description of inflation.

\item[$dS_{s}$] Finally, our variables also formally permit to satisfy the fixed point condition \eqref{dS fixed point condition} at the limit $f(\phi_s)\rightarrow -1$ where the effective gravitational constant \eqref{eq:G_eff} becomes singular. At this value $\phi_s$, the effective potential \eqref{eq:V_eff} turns singular as well and the situation is in some ways similar to $dS_{\infty}$, i.e.\ de Sitter like accelerated expansion with $w_{\rm eff}=-1$, but $H\rightarrow \infty$ while $z=\tfrac{\dot{\phi}}{H} \rightarrow 0$. Of course, this point is only available in the nonminimally coupled case. Numerical checks show this point behaves as a saddle, consistent with the property that the singular value of the effective potential can only be a maximum according to our assumptions. Although the variation of the effective gravitational constant is rather restricted by observations in the late universe, the possibility of this regime as a  rallying point for inflation can be entertained. Like $dS_\infty$, the $dS_s$ point does not satisfy the condition \eqref{eq:GR_limit_V} of GR limit either.

\end{enumerate} 

The determination of the eigenvectors associated with each critical point suggests that scalar-tensor theories are able to explain the thermal history of the universe, i.e.\ going from a scalar field dominated inflationary expansion to radiation dominated era to matter dominated era to the final scalar field dominated accelerated expansion (dark energy) epoch. This sequence can be realized generically in a model where the corresponding fixed points exist as three saddles and an attractor, respectively. These fixed points shape the flow of trajectories and ensure that the passing by solutions exhibit the expansion rate appropriate for the relevant era.
While the radiation domination, matter domination, and dark energy domination points are generated quite easily, not every STG model provides de Sitter point suitable for inflation, i.e.\ a saddle where the repulsive direction guides the trajectories away, so that the early accelerated expansion can end. Still, in many models de Sitter saddle is available, either as a regular local maximum ($dS$), an asymptotic maximum ($dS_\infty$), or a singular value ($dS_s$) of the effective potential.

\section{Example: general relativity with cosmological constant}
\label{sec:example_gr}

To get a glimpse of how the physics is captured in our dynamical system, let us begin by considering the simple case of GR with radiation and matter, while taking $\phi\equiv constant, f(\phi)\equiv0, V(\phi)\equiv \Lambda$. 
In this case, only the variables $x$ and $y$ are dynamical and the system \eqref{dynsys x}-\eqref{dynsys phi} reduces to
\begin{eqnarray}
\label{eq:GR x}
x' &=& 4 x^2 + 3 xy - 4 x\,, \\
\label{eq:GR y}
y' &=& 3 y^2 + 4 xy - 3 y \,.
\end{eqnarray}
Assuming the cosmological constant is positive, $\Lambda>0$, and denoting $\Omega_\Lambda=\frac{\Lambda}{3 H^2}$, the Friedmann constraint 
\begin{equation}
1 = x + y + \Omega_\Lambda \,,
\end{equation}
implies that the two dimensional phase space is bounded, delimited by 
\begin{align}
\Psi_{\Lambda}=\left\lbrace (x,y) \in \mathbb{R}^2 \Big|   0 \leq x, y <1;\,\, 0\leq x+y < 1 \right\rbrace\,.
\end{align} 
The same system has been presented before in the literature \cite{Fay:2014fta,Bahamonde:2017ize}, and here we invoke it to guide the reader into more involved cases which follow.

\begin{figure}
	\centering
	\subfigure[]{
		\includegraphics[width=5cm, height=5cm]{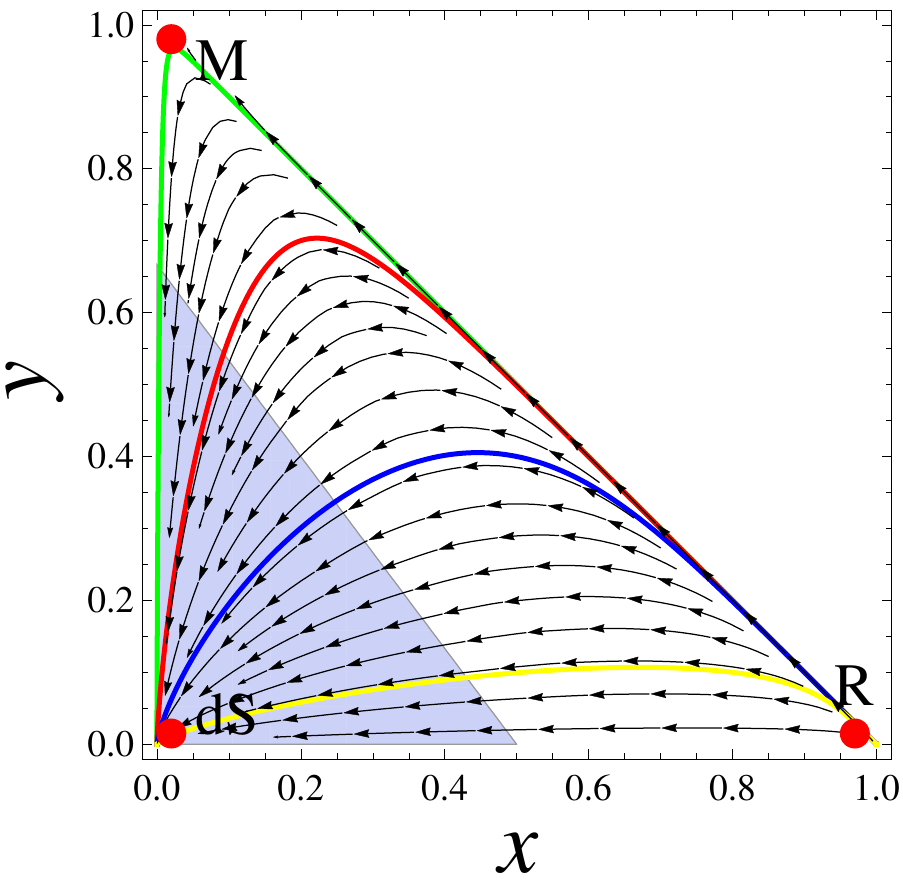} \label{GR_phase_space_fig}}
	\qquad
	\subfigure[]{
		\includegraphics[width=5cm, height=5cm]{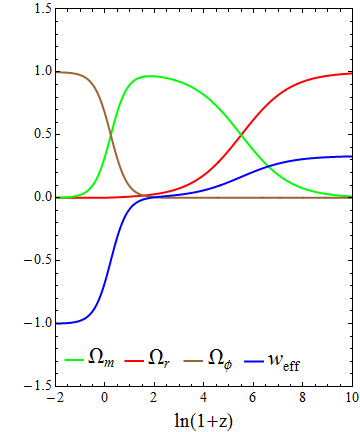} \label{GR_parameters} }
	\caption{(a) Cosmological phase space in GR with radiation, dust matter, and cosmological constant. The shaded area represents the region of accelerated expansion. (b)  Evolution of  dust matter, radiation, and scalar field energy density parameters along with the effective equation of state along the solution represented by the green trajectory on panel (a).}
\end{figure}

From the system \eqref{eq:GR x}-\eqref{eq:GR y}, it is easy to see that there are three fixed points situated at the boundary of the phase space, characterized by their properties as follows:
\begin{itemize}
\item[$R$] at $(1, 0)$ is a repeller and corresponds to radiation dominated universe, $\Omega_r=1$, $w_{\mathrm{eff}}=\tfrac{1}{3}$;
\item[$M$] at $(0, 1)$ is a saddle and corresponds to dust matter dominated universe, $\Omega_m=1$, $w_{\mathrm{eff}}=0$;
\item[$dS$] at $(0, 0)$ is an attractor and corresponds to de Sitter universe, $\Omega_\Lambda=1$, $H^2=\frac{\Lambda}{3}$, $w_{\mathrm{eff}}=-1$.
\end{itemize}
The properties of the fixed points can be inferred from the eigenvalues of the linearized system at the fixed points, as a particular subcase in the discussion of the previous section, but are also directly evident from the phase diagram on Fig.\ \ref{GR_phase_space_fig}.
It should be stressed that the solutions sitting exactly at the fixed points are relatively uninteresting, corresponding here to universes filled with only radiation, dust matter, and cosmological constant, respectively. The importance of fixed points, however, lies in how they determine the behavior of solutions around them: repellers act like sources as nearby trajectories depart from them, attractors act like sinks as nearby trajectories arrive at them, while the saddles repel in one direction and attract in another direction. On Fig.\ \ref{GR_phase_space_fig}, we see how the generic trajectories start near the radiation domination fixed point (R), evolve towards the matter domination fixed point (M), but turn away from it and end up at the de Sitter point (dS). At the boundaries, there are also three special (``heteroclinic'') trajectories, one starting at R and running directly into M, another starting at R and running into dS, and third starting at M and running into dS. These correspond to universes without cosmological constant, matter, and radiation, respectively. Technically speaking, the points R and M, as well as the orbit connecting them are not part of the phase space $\Psi_\Lambda$ if $\Lambda>0$, but to understand the physics it is certainly illuminating to include them. The de Sitter point $dS$ is a global attractor as it lies in the intersection of invariant submanifolds $x=0$ and $y=0$.

The behavior of generic trajectories in the phase space is in qualitative agreement with the history of our observable universe. Earlier the universe was dominated by radiation, then as radiation dilutes the dust matter takes the role of the leading component for a certain period, while in the future both radiation and dust matter densities drop and the reign of the cosmological constant begins. At some point before the complete domination of the cosmological constant, the universe starts to expand in an accelerated fashion, as the trajectories enter into $ w_{\rm eff}<-\tfrac{1}{3}$ regime, marked by the shaded area on Fig.\ \ref{GR_phase_space_fig}. On the figure, the late history of our observable universe is best represented by the green trajectory which comes very close to the M point. We can observe the evolution of the relative densities of radiation, matter, and cosmological constant, as well as the effective barotropic index, along this trajectory on Fig.\ \ref{GR_parameters}. (Here we employ a useful convention to follow the evolution in terms of the logarithm of redshift, related to the scale factor by $1+z=\tfrac{1}{a}$, whereby the present scale factor is normalized, $a(t_{\rm now})=1$. In this way, different epochs can be conveniently and visibly captured on a single plot, as in e.g.\ Refs.\ \cite{Dutta:2017fjw,Sami:2012uh}.)
The majority of other trajectories in the phase space correspond to larger values of the cosmological constant, whereby the matter component in the universe has less chance to dominate.

The system \eqref{eq:GR x}-\eqref{eq:GR y} gave us a correct qualitative picture of the latter stages of the evolution of the universe, owing to the properties of the fixed points. What is missing is the epoch of early accelerated expansion, aka inflation. Although there is an accelerating region in the phase space surrounding the dS fixed point, it is unable to provide early time acceleration, since all the trajectories only enter it, while none goes out from this patch. This follows from the attractive nature of the dS fixed point, which draws all trajectories into its fold and lets none to escape. For an early transient period of accelerated expansion, we need another fixed point which is of the dS type, but has the characteristics of a repeller or at least a saddle.

\section{Example: Minimally coupled scalar field with quartic potential}
\label{sec:example_minimal}

As the next step, let us consider minimally coupled scalar with a quartic potential,
\begin{equation}
f=0 \,, \qquad
V=V_0 \phi^4 +\Lambda\,,
\end{equation}
where $V_0>0$, $\Lambda>0$ as an example. The corresponding dynamical system, although cast in different variables was considered before in the case of absent matter fluids in Ref.\ \cite{Skugoreva:2014gka} and with a single fluid but without cosmological constant in Ref.\ \cite{Alho:2015cza}.

Compared to the previous section, including a dynamical scalar field makes the phase space four dimensional. The positivity of the potential imposes a constraint via the Friedmann equation \eqref{constraint_eqn},
\begin{equation}
\label{eq:Friedmann_constraint_min_cpld}
z^2 < 6 - 6x -6y \,.
\end{equation}
From the notes in Sec.\ \ref{sec:fixed_points}, we can easily find out the fixed points: 
\begin{itemize}
\item[$R$] at $(1,0,0,\phi)$ form a line along the $\phi$ direction, all these points are saddles in character.
\item[$M$] at $(0,1,0,\phi)$ also form a line along the $\phi$ direction, and are saddles as well.
\item[$dS_0$] at $(0,0,0,0)$ corresponds to de Sitter universe with $H^2=\frac{\Lambda}{3}$. Using the results of appendix \ref{CMTdS1},  we can determine it to be a stable node. 
\item[$dS_\infty^\pm$] at $(0,0,0,\pm\phi_*)$ lie in the asymptotics $\phi_*\rightarrow \infty$ and are saddles. Here we have de Sitter like equation of state, $w_{\rm eff}=-1$, but the Hubble parameter diverges, $H\rightarrow \infty$.
\end{itemize}

\begin{figure}
\centering
\subfigure[]{\includegraphics[width=5cm, height=5cm]{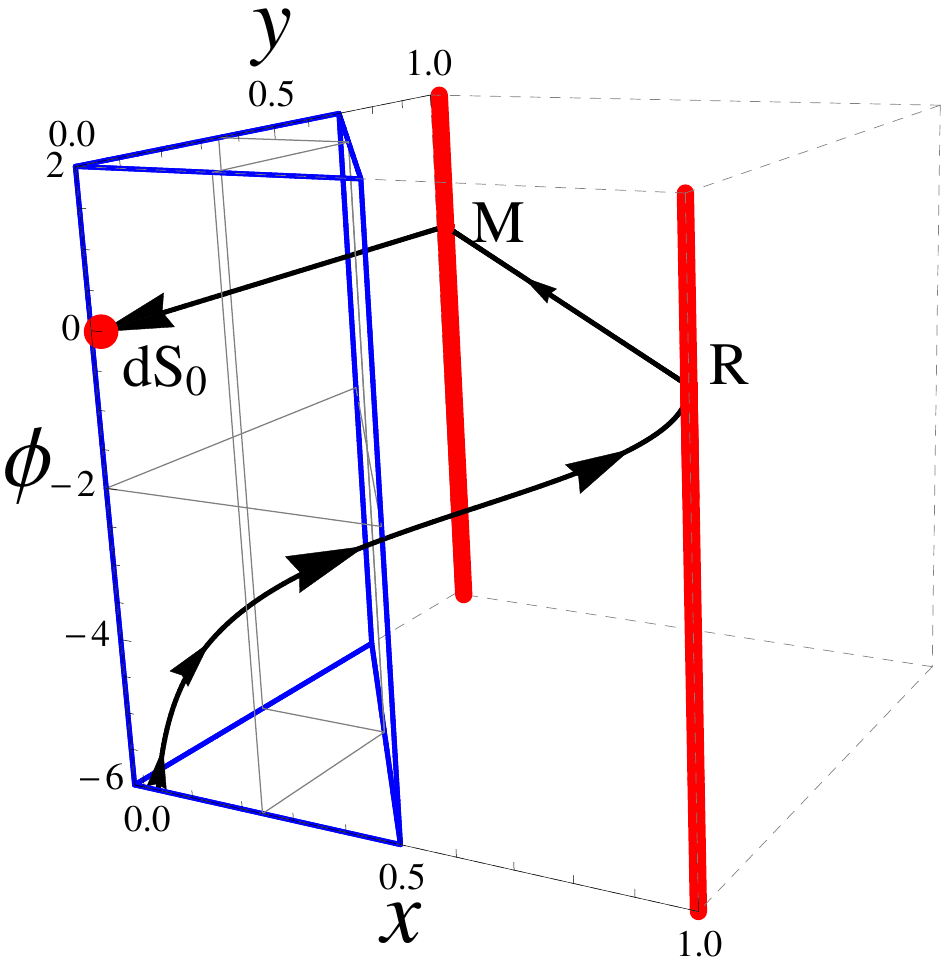} \label{3D_x_y_phi_min_cpld}}
	\qquad
\subfigure[]{\includegraphics[width=5cm, height=5cm]{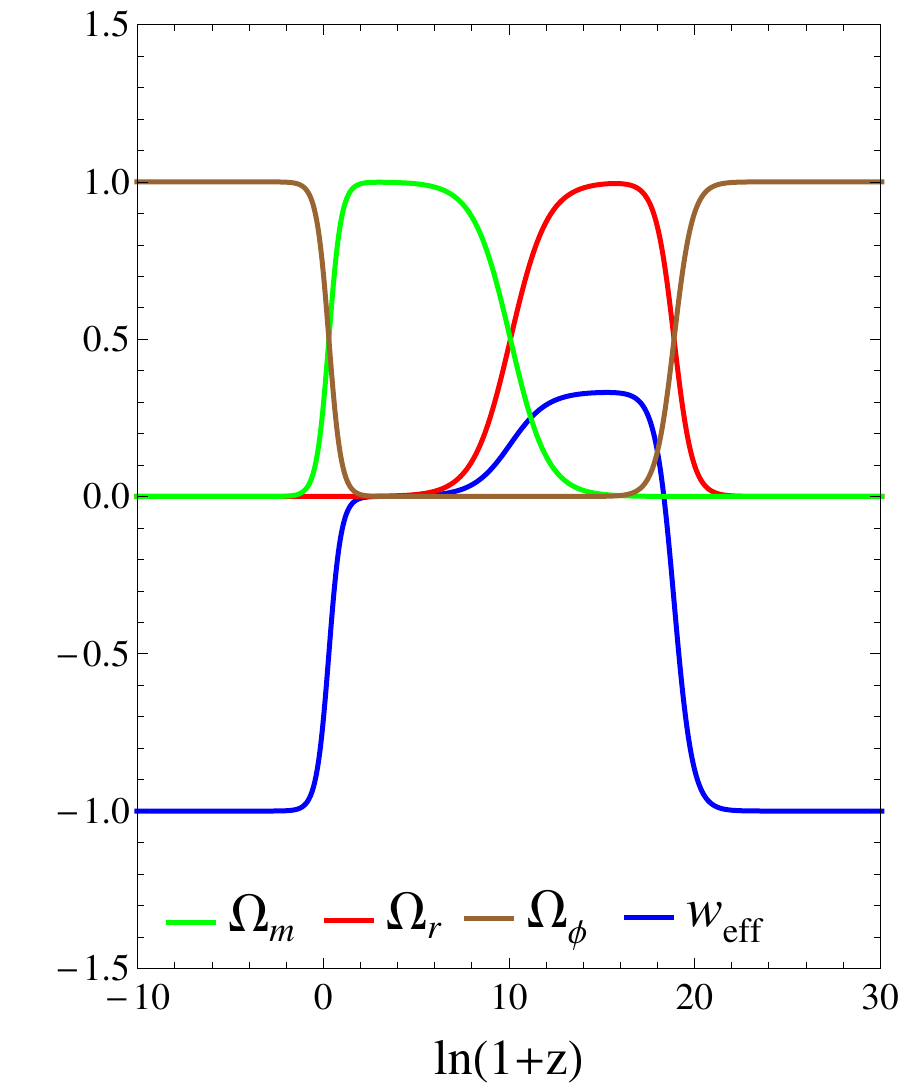} \label{parameters_min_cpld_new} }
	\\
\subfigure[]{\includegraphics[width=5cm, height=5cm]{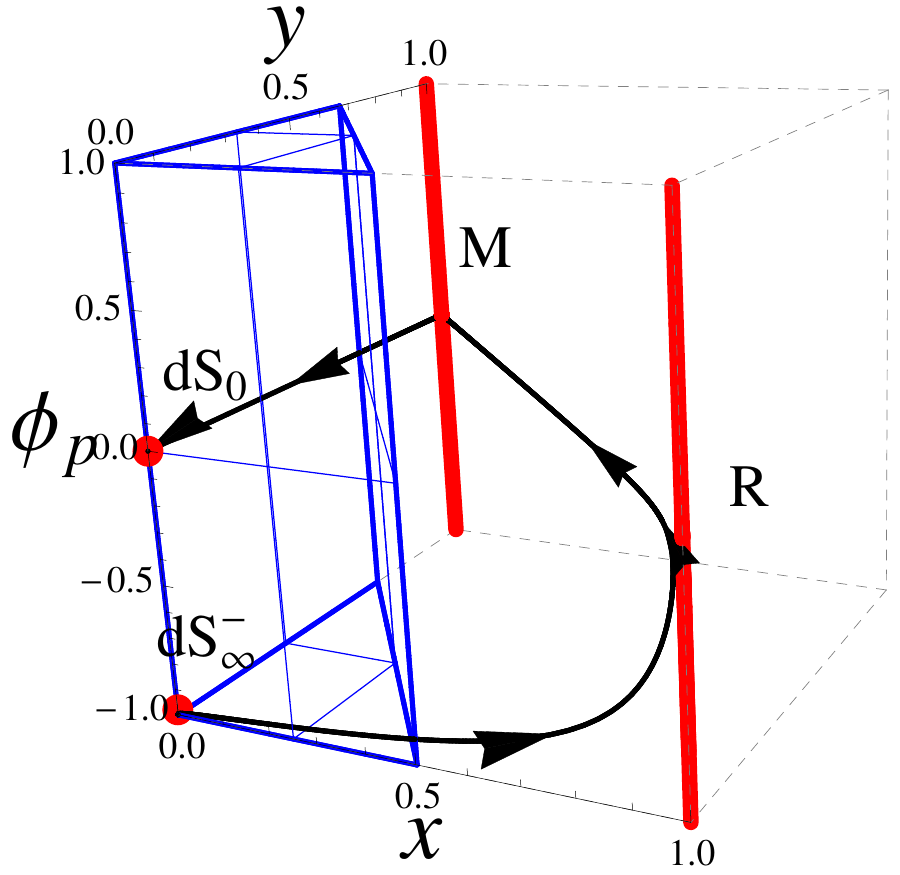} \label{poincare_3D_min_cpld_neg} }	
     \qquad
\subfigure[]{\includegraphics[width=5cm, height=5cm]{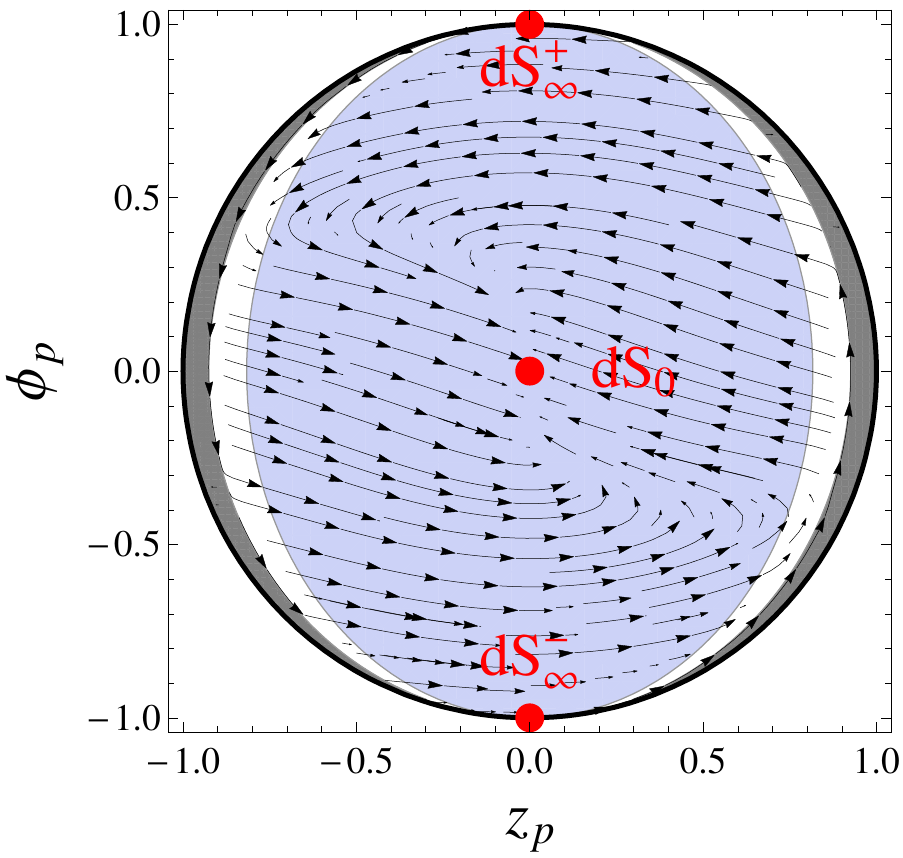} \label{poincare_2D_min_cpld}}
\caption{Dynamics and phase space for the system \eqref{dynsys x}-\eqref{dynsys phi} in the minimally coupled case, $f(\phi)=0$, $V(\phi)=V_0 \phi^4+\Lambda$ with $V_0=1$, $\Lambda=1$. (a) Phase trajectory in the $(x,y,\phi)$ dimensions, the region enclosed by a grid represents the accelerated expansion. (b) Evolution of dust matter,  radiation,  and scalar field energy density parameters along with the effective equation of state along the depicted solution. (c) Phase trajectory in the $(x,y,\phi_p)$ dimensions, where the range of the scalar field is compactified to $-1\leq \phi_p \leq 1$. (d) Projection of the phase flow of the fully compactified system at the $x=y=0$ plane, the blue area represents the region of accelerated expansion while the dark gray marks an area ruled out by the Friedmann constraint \eqref{eq:Friedmann_constraint_min_cpld}. }  
\end{figure}

The behavior of one representative phase trajectory in a finite patch of ($x$, $y$, $\phi$) subspace is depicted in Fig. \ref{3D_x_y_phi_min_cpld}, where the region of accelerated expansion is indicated by the blue contour and the fixed points shown in red. It can be seen that the trajectory evolves from the accelerating region towards a radiation dominated point $R$, then to a matter dominated point $M$, and eventually returns to the accelerated region and settles down approaching the  de Sitter point $dS_0$. Recall the result of Sec.~\ref{sec:fixed_points} that the attractive direction of the radiation domination point was lying in the $y=0$ plane where dust matter density vanishes. On the plot \ref{3D_x_y_phi_min_cpld} this reflects in the fact that the trajectory approaches $R$ by being close to the $y=0$ plane. When the universe evolves from inflation to radiation domination the matter density stays low, it will increase only after passing by the radiation domination point. Analogously, the repulsive direction of $M$ lies in the $x=0$ plane, thus after the matter domination,  radiation density has no tendency to increase again. We just notice how such basic thermodynamic features are encoded in the phase space properties.

The time evolution of energy density parameters along with the effective equation of state for this solution can be read off from Fig. \ref{parameters_min_cpld_new}. Here, we see how the universe evolves from the scalar field dominated inflationary era ($\Omega_{\rm  de} \approx 1$, $w_{\rm eff} \approx -1$), passes on to the radiation domination epoch ($\Omega_{\rm  r} \approx 1$, $w_{\rm eff} \approx \frac{1}{3}$), followed by the matter domination epoch  ($\Omega_{\rm  m} \approx 1$, $w_{\rm eff} \approx 0$), and eventually settles in an accelerated expansion regime dominated by the scalar field as dark energy ($\Omega_{\rm  de} \approx 1$, $w_{\rm eff} \approx -1$). Although the history is presented here on a logarithmic scale, we can still notice how the fixed points are instrumental in making each of the epochs pronounced enough. Indeed, near the fixed points the flow along the trajectory (the evolution of the solution measured in time, scale factor, or redshift) ``slows down'' as the ``force'' (r.h.s.\ of the system \eqref{dynsys x}-\eqref{dynsys phi}) becomes small and hence the derivatives of the dynamical variables become small too. On the other hand, the transition from one epoch to another occurs relatively ``quickly'', since while the trajectory passes from the vicinity of one fixed point to another, the ``force'' is bigger in the middle regions and the dynamical variables evolve ``faster''.

The finite phase space \ref{3D_x_y_phi_min_cpld} does not reveal the earlier path of the trajectory, where does it originate from. Thus let us check a hybrid plot \ref{poincare_3D_min_cpld_neg}, where the variables $z, \phi$ are made compact in analogy with \eqref{poinc}, i.e.\ mapped to the range $(-1,1)$. Tracing the same trajectory backwards 
takes us into the vicinity of the asymptotic fixed point $dS_\infty^-$. Therefore this point, as well as its cousin $dS_\infty^+$ are indeed responsible for inflation, because around these points the trajectories can exhibit de Sitter like expansion ($w_{\rm eff} \approx -1$) which lasts sufficiently long.
Since this point is a saddle, there is only one trajectory that exactly starts at this point. It is the one that flows out of $dS_\infty$ along the repulsive eigendirection. All other trajectories in the phase neighbourhood come from somewhere else (other asymptotic regions), but closely approach this particular solution as they evolve, since all other eigendirections are attractive. In principle,  we could follow our example trajectory further back in time, but as we already are in the asymptotics  the numerical errors mount. The main feature is that inflation is a generic phenomenon for a wide range of initial conditions.

For a true glimpse of the global dynamics of the system, we should compactify the full phase space by the Poincar\'e transformation \eqref{comp_phase}. The asymptotic fixed points reside at the boundary of the compact space, $dS_\infty^+ (0,0,0,1)$ and  $dS_\infty^- (0,0,0,-1)$. Fig.~\ref{poincare_2D_min_cpld} depicts the $x=y=0$ slice through the 4-dimensional compact phase space, where points on the circumference of the circle represent the asymptotic infinity. The gray zone is the region prohibited by the Friedmann constraint \eqref{eq:Friedmann_constraint_min_cpld}, and the blue shaded area stands for the region of accelerated expansion. 
We have also plotted a projection of the direction of the trajectories' flow, as they pass through or along this phase space slice.
By closely observing the phase flow and also doing quick numerical investigations, we can recognize that the points $dS_\infty^+$ and  $dS_\infty^-$ behave as saddle points, since the trajectories come close to them and then evolve away, eventually heading towards the attractor de Sitter point ($dS_0$). The logic of the phase flow also implies that the $dS_\infty^\pm$ and $dS_0$ points are connected by a heteroclinic orbit lying exactly on the $x=y=0$ plane. We do not see the $R$ and $M$ points here because they reside in the dimensions that are not shown on this plot. However, before reaching the global attractor $dS_0$,  the majority of the trajectories also wander in the $x$ and $y$ dimensions and pay a visit to these points as well. 
Putting all this information together in a global picture, we conclude that the basic cosmic history of inflation, radiation domination, matter domination, and final dark energy occurs as a rather generic property for the trajectories of the model, guaranteed by the suitable structure of fixed points.

\begin{figure}
        \centering
        \subfigure[]{%
            \includegraphics[width=5cm,height=3.5cm]{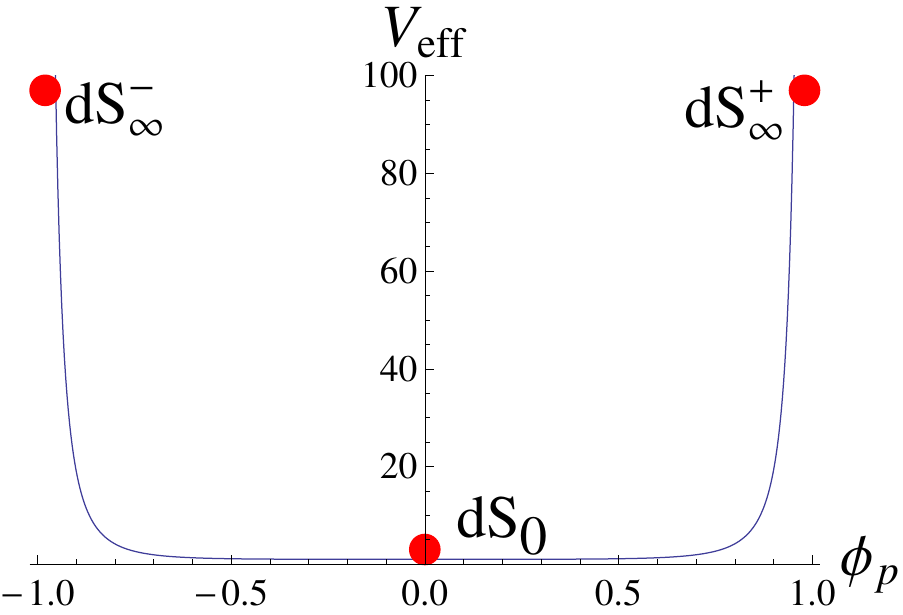}\label{V_eff_xi_zero}}
        \qquad
        \subfigure[]{%
            \includegraphics[width=5cm,height=3.5cm]{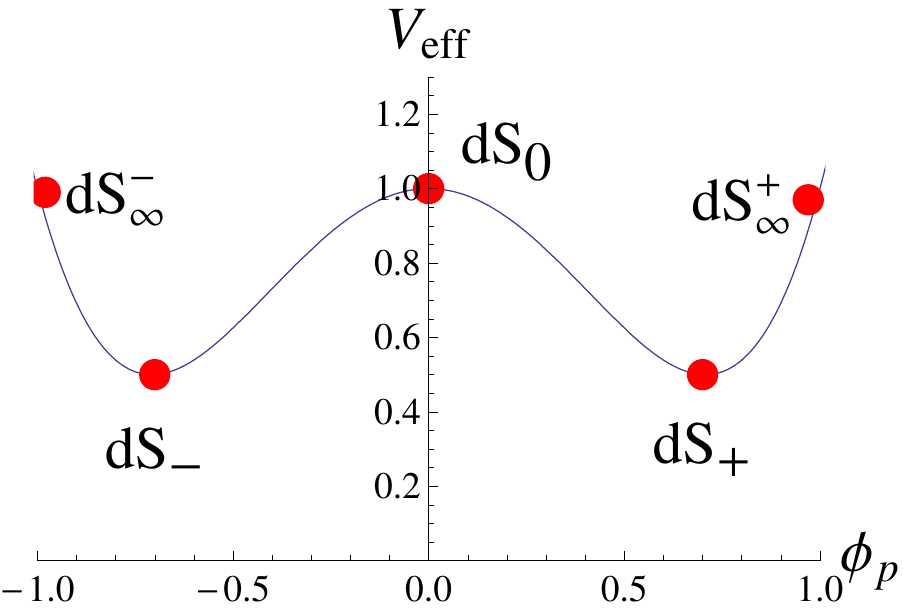}\label{V_eff_xi_pos}}
            \qquad
        \subfigure[]{%
            \includegraphics[width=5cm,height=3.5cm]{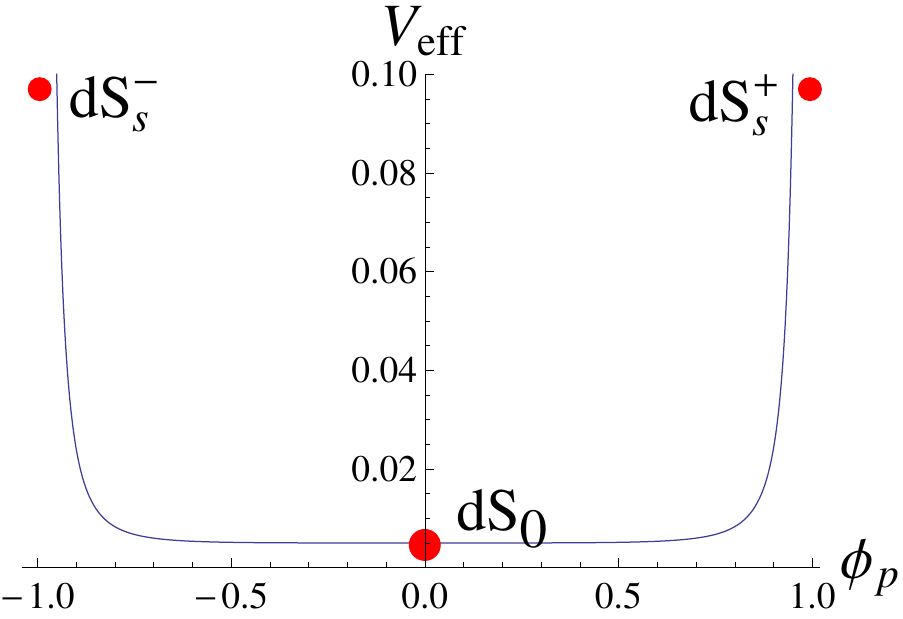}\label{V_eff_xi_neg}}
            \caption{The plot of the effective potential $V_{\rm eff}$ for (a) minimal coupling $(V_0=1, \Lambda=1, \xi=0)$, (b) positive coupling $(V_0=1, \Lambda=1, \xi=1)$, and (c) negative coupling $(V_0=0.001, \Lambda=0.005, \xi=-0.005)$.}
\label{V_eff_plot}
\end{figure} 


\begin{figure}[h!]
\centering
\subfigure[]{\includegraphics[width=5cm,height=5cm]{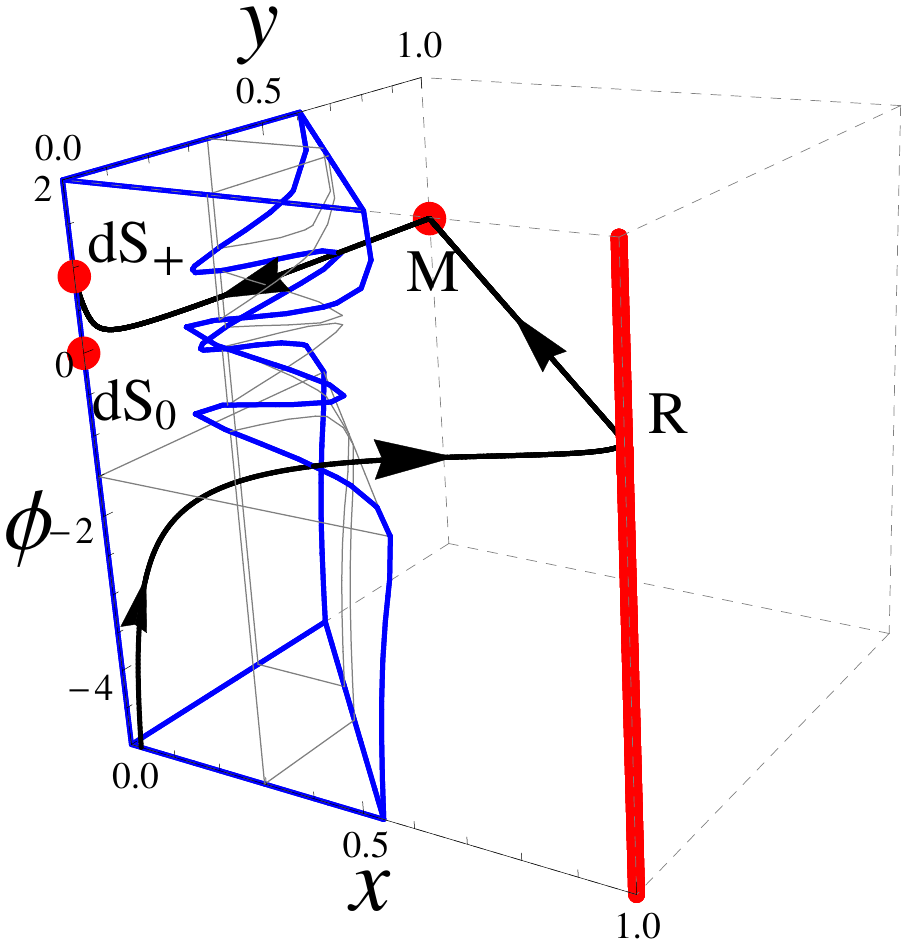}\label{3D_x_y_phi_phi2_pow_pot}}
    \qquad
\subfigure[]{\includegraphics[width=5cm,height=5cm]{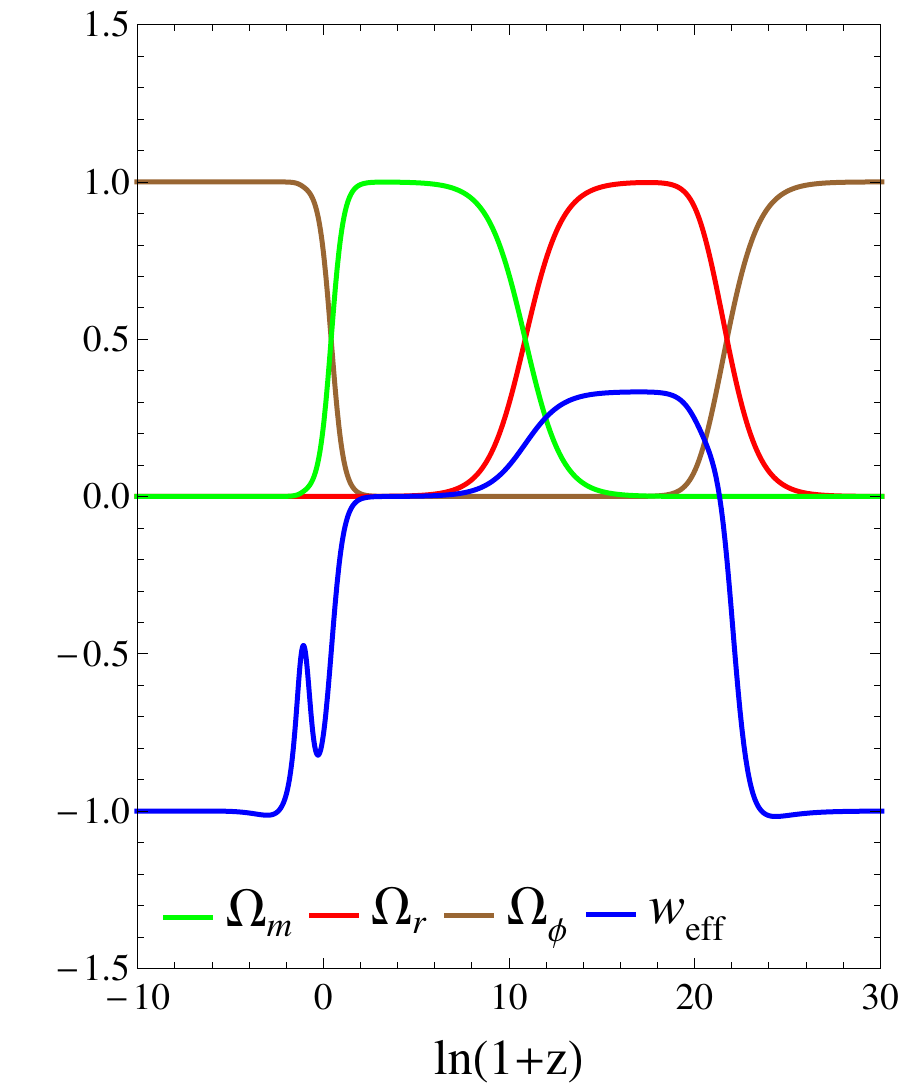}\label{parameters_phi2_pow_pot}}  
	\\
\subfigure[]{\includegraphics[width=5cm,height=5cm]{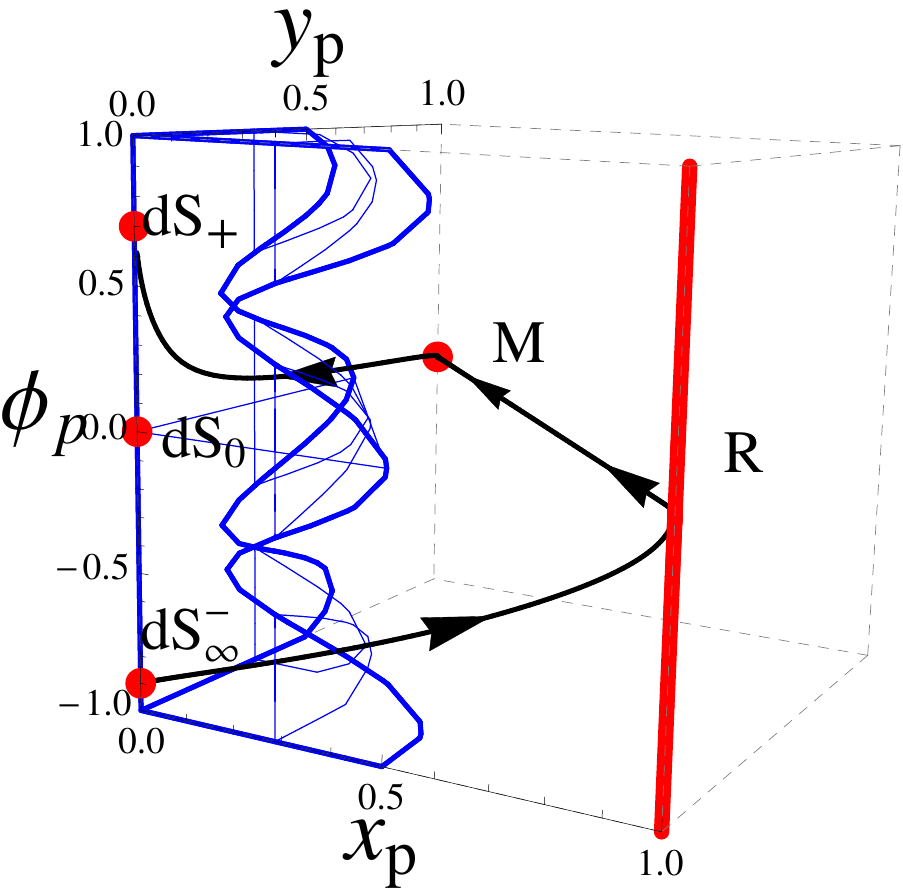}\label{poincare_3D}}
	\qquad 
\subfigure[]{\includegraphics[width=5cm,height=5cm]{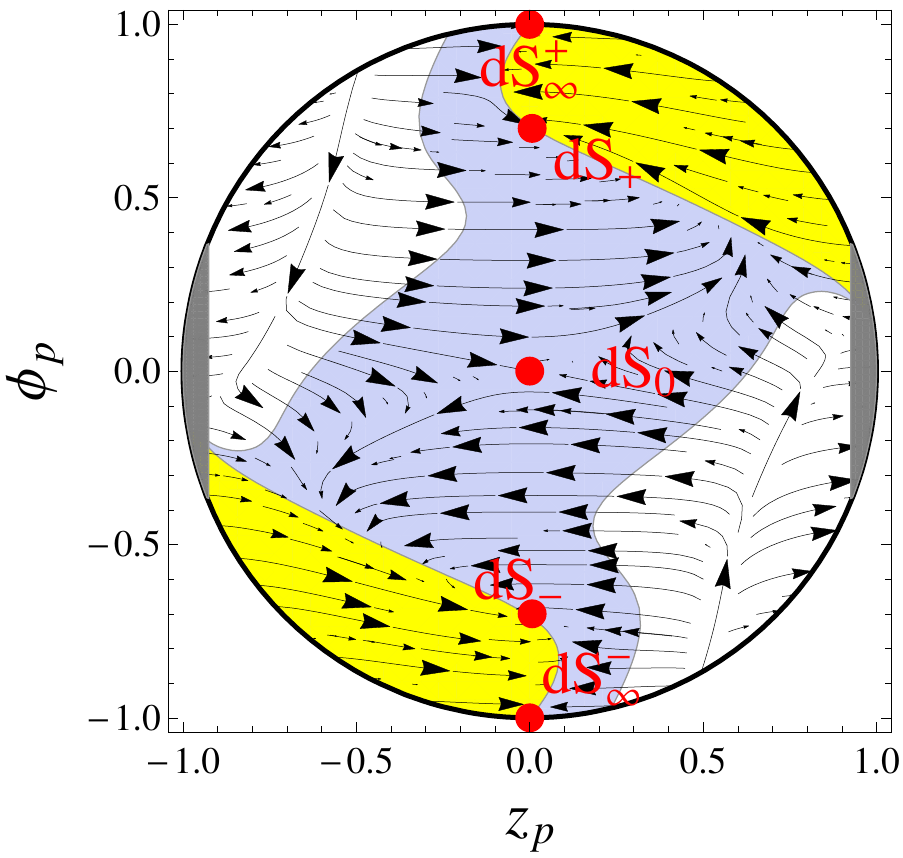}\label{poincare_2D}}
	\\
\subfigure[]{\includegraphics[width=5cm,height=5cm]{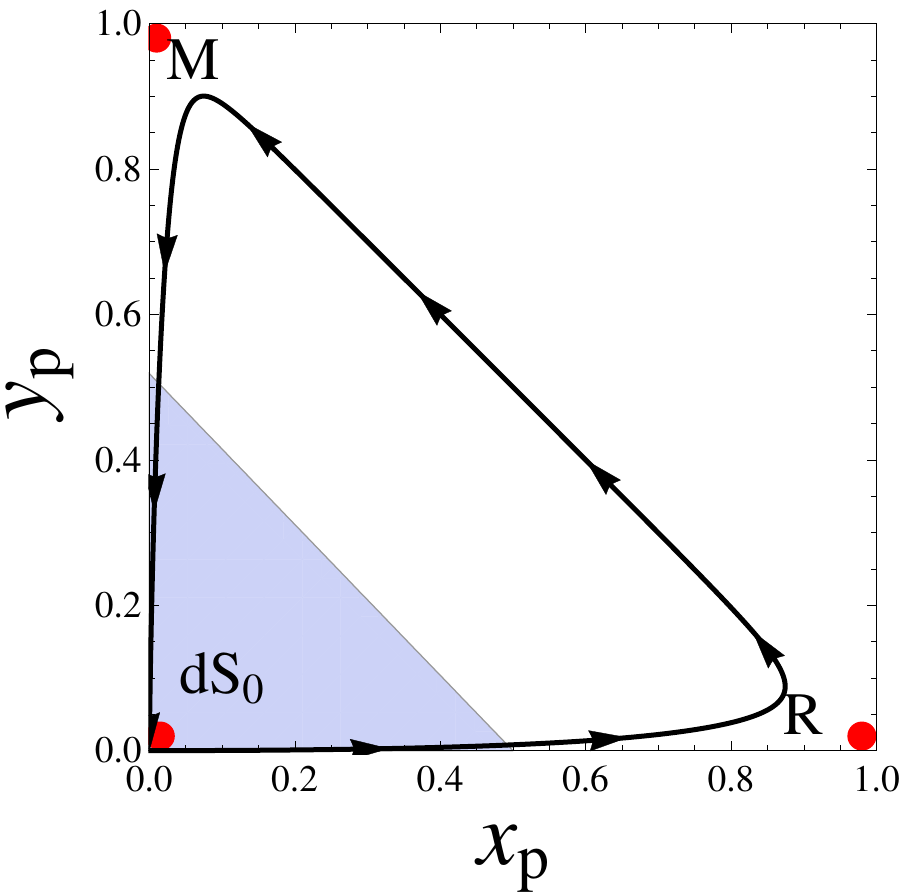}\label{acc_2d}}
             \qquad
\subfigure[]{\includegraphics[width=5cm,height=5cm]{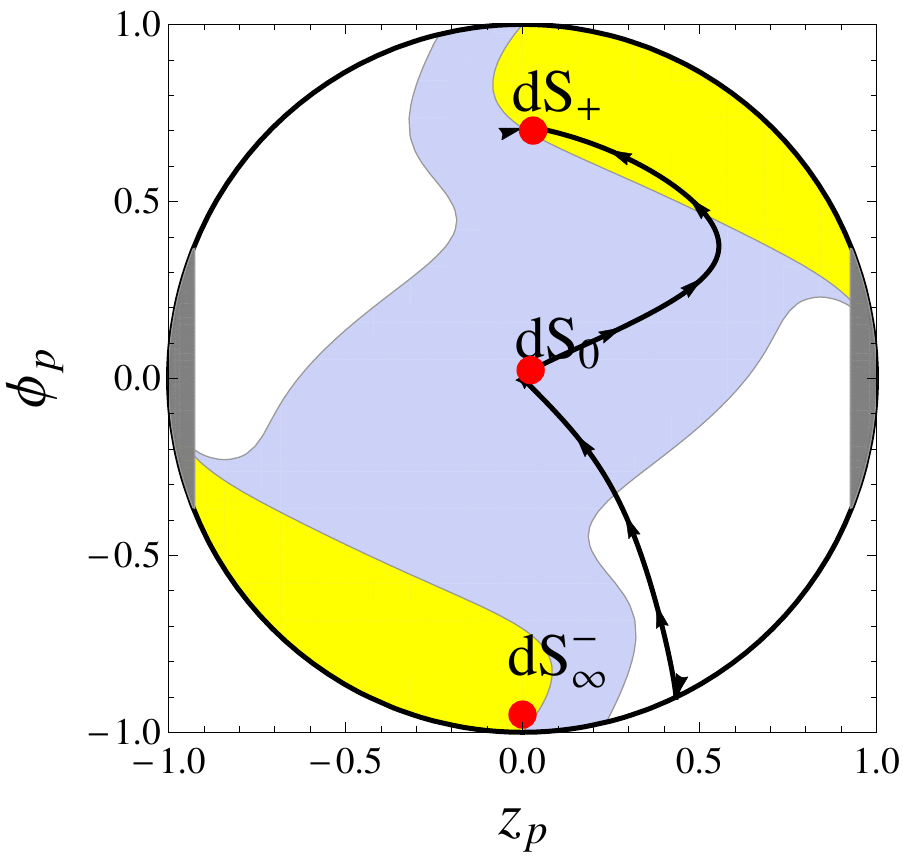}\label{2d_z_phi_pcre}}
\caption{The dynamics in the system \eqref{dynsys x}-\eqref{dynsys phi} in the nonminimal quadratic coupling, quartic potential case \eqref{eq:nonminimal_quartic_potential} for an example of positive coupling ($\xi=1, V_0=1, \Lambda=1$). (a) An example phase trajectory in the $(x,y,\phi)$ dimensions, the region enclosed by a grid represents the accelerated expansion. (b) Evolution of dust matter, radiation,  and scalar field energy density parameters along with the effective equation of state along the depicted solution. (c) The same phase trajectory in the $(x_p,y_p,\phi_p)$ dimensions, where the range of the scalar field is compactified to $-1 \leq \phi_p \leq 1$. (d) Projection of the phase flow of the fully compactified system at the $x_p=y_p=0$ plane. Here and on the following subplots the area shaded in blue represents the region of accelerated expansion, the area in yellow represents the superaccelerated expansion, and the area in dark gray is ruled out by the Friedmann constraint \eqref{eq:Friedmann_constraint_nonmin_cpld}.  (e) Projection of the example trajectory to the $z_p=\phi_p=0$ plane. (f) Projection of the example trajectory to the $x_p=y_p=0$ plane. }
\label{fig:xi_positive}
\end{figure}
\section{Example: Quadratic nonminimal coupling, quartic potential}
\label{sec:example_nonminimal}

As the main example, let us consider the case of quadratic nonminimal coupling and quartic potential with an additional constant term,
\begin{equation}
\label{eq:nonminimal_quartic_potential}
f(\phi)=\xi \phi^2, \qquad V(\phi)=V_0 \phi^4 +\Lambda \,.
\end{equation}
It can be considered as a simplified version of the nonminimally coupled Higgs model \cite{Bezrukov:2007ep} as the quartic piece dominates at large field values and is responsible for early universe inflation, while the constant piece dominates late universe as dark energy. The model is also interesting because it allows us to witness attractive and repulsive regular $dS$ fixed points as well as asymptotic $dS_\infty$ and singular $dS_s$ fixed points, described in Sec.\ \ref{sec:fixed_points}. Without the radiation and dust matter components, the model \eqref{eq:nonminimal_quartic_potential} was analyzed before in Ref.\ \cite{Skugoreva:2014gka} in different variables. The predictions for inflationary observables can be found in the literature \cite{Kaiser:1994vs, Okada:2010jf, Bostan:2019fvk} or quickly computed from the generic algorithm \cite{Jarv:2016sow}.

The phase space is still 4-dimensional with a few extra features and novel fixed points compared to the minimally coupled case. The positivity of the potential imposes a constraint via the Friedmann equation \eqref{constraint_eqn},
\begin{equation}
\label{eq:Friedmann_constraint_nonmin_cpld}
\frac{z^2}{1+\xi\phi^2} - \frac{12 \xi z \phi}{1+\xi \phi^2} < 6-6x-6y \,.
\end{equation}
In the case of $\xi<0$ the positivity of the effective gravitational constant \eqref{eq:G_eff} also sets a limit
\begin{equation}
\label{eq:Geff_constraint_nonmin_cpld}
\phi^2< -\frac{1}{\xi} \,, 
\end{equation} 
while that boundary is marked by the effective potential \eqref{eq:V_eff} becoming singular.
 
Recalling the general results of Sec. \ref{sec:fixed_points}, the fixed points $R$ and $M$ are independent of the model parameters, but the set and character of de Sitter points is determined by the conditions \eqref{dS fixed point condition}, \eqref{dS fixed point eigenvalue positive} 
and depends on the value of the nonminimal coupling parameter $\xi$, most easily to understood in reference to the shape of the effective potential \eqref{eq:V_eff} plotted on Fig.\ \ref{V_eff_plot}. In summary, the fixed points are the following:
\begin{itemize}
\item[$R$] at $(1,0,0,\phi)$ form a line along the $\phi$ direction, all these points are saddles in character.
\item[$M$] at $(0,1,0,0)$ is just one point (not a line as in the minimally coupled case) matching the GR limit \eqref{eq:GR_limit_M}, it is a saddle.
\item[$dS_0$] at $(0,0,0,0)$ is a saddle for $\xi>0$ when it resides at the local maximum of the effective potential, and an attractor for $\xi<0$ when it sits at the local minimum of the effective potential. In the latter case it is a stable node if $-\frac{3}{16}<\xi<0$ and stable focus if $\xi<-\frac{3}{16}$. It corresponds to the GR limit \eqref{eq:GR_limit_V}.
\item[$dS_\pm$] at $(0,0,0,\pm \sqrt{\tfrac{\Lambda\xi}{V_0}})$ exist only for $\xi>0$ and are attractors occupying the local minimimum of the effective potential. More precisely, their nature is stable focus for all values of $\xi$. They also satisfy the GR limit \eqref{eq:GR_limit_V}.
\item[$dS_\infty^\pm$] at $(0,0,0,\pm\phi_*)$ are situated in the asymptotics $\phi_*\rightarrow \infty$. These points have a meaning only for $\xi>0$ when they coincide with the asymptotic maximum of the effective potential and correspond to saddles. (For $\xi<0$ the asymptotic region is unphysical.) 
\item[$dS_s^\pm$] at $\left(0,0,0,\pm\frac{1}{\sqrt{-\xi}}\right)$ exist only in the case of negative nonminimal coupling $\xi<0$. These points are saddles and correspond to a singularity of the effective potential.
\end{itemize}

Note that the emergence of $dS_\pm$ happened by the virtue of adding a positive constant $\Lambda$ to the potential. Without it, the system would have had only one regular de Sitter fixed point $dS_0$ at the  minimum of the effective potential. Equipped with the general knowledge of the phase space and fixed points, let us illustrate the cosmological dynamics and follow a particular trajectory in the phase space, treating the $\xi>0$ and $\xi<0$ cases separately.

\begin{figure}
\centering
\subfigure[]{\includegraphics[width=5cm,height=4cm]{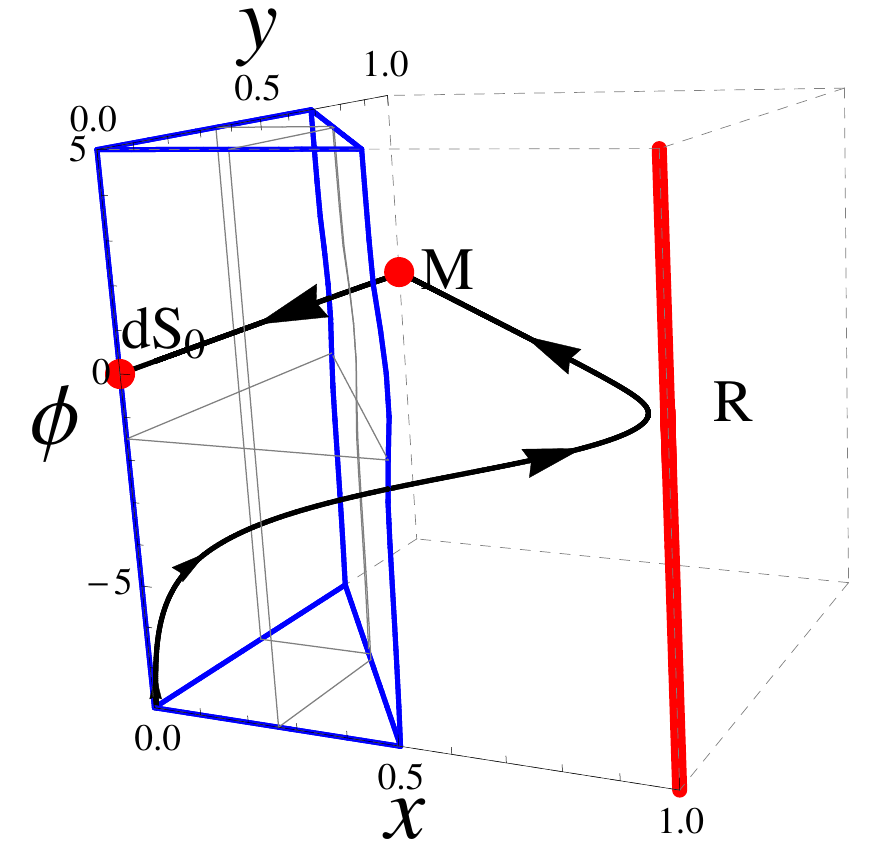}\label{3D_x_y_phi_phi2_pow_pot_neg}}
    \qquad
\subfigure[]{\includegraphics[width=5cm,height=4cm]{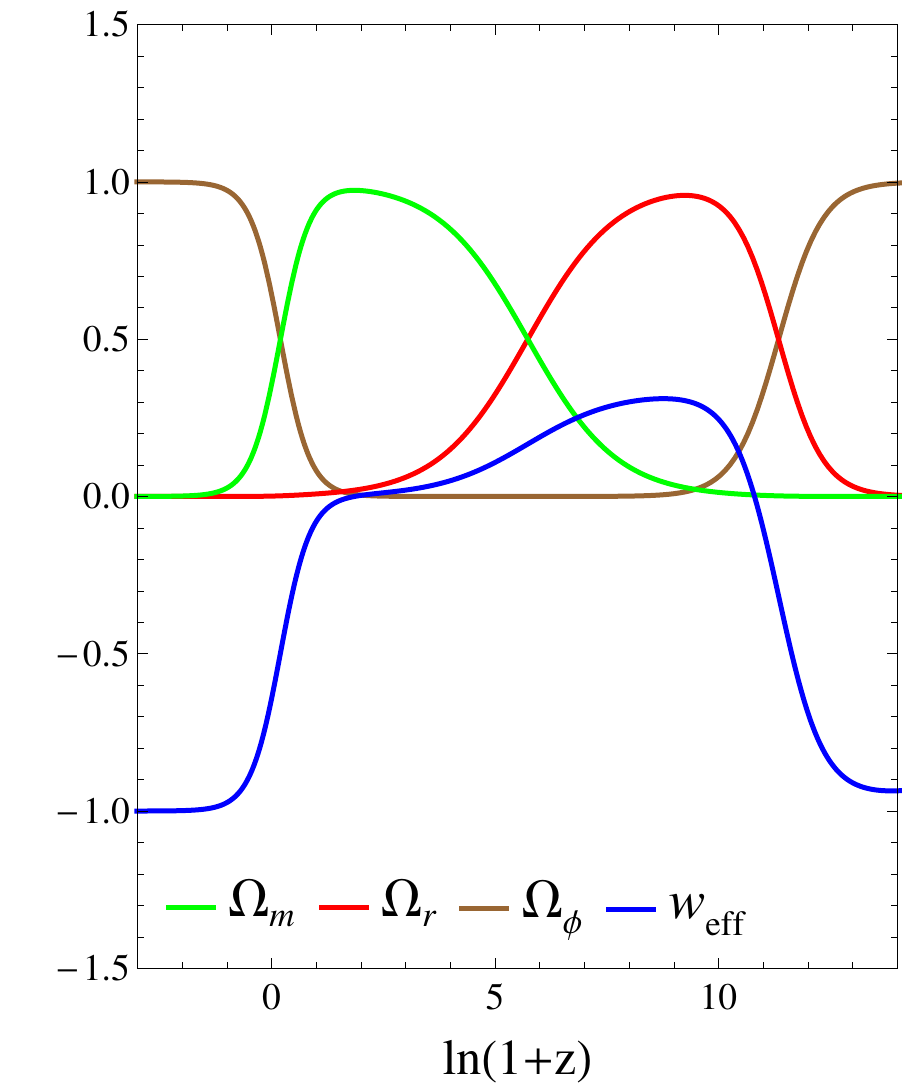}\label{parameters_phi2_pow_pot_neg}}  
	\\
\subfigure[]{\includegraphics[width=5cm,height=5cm]{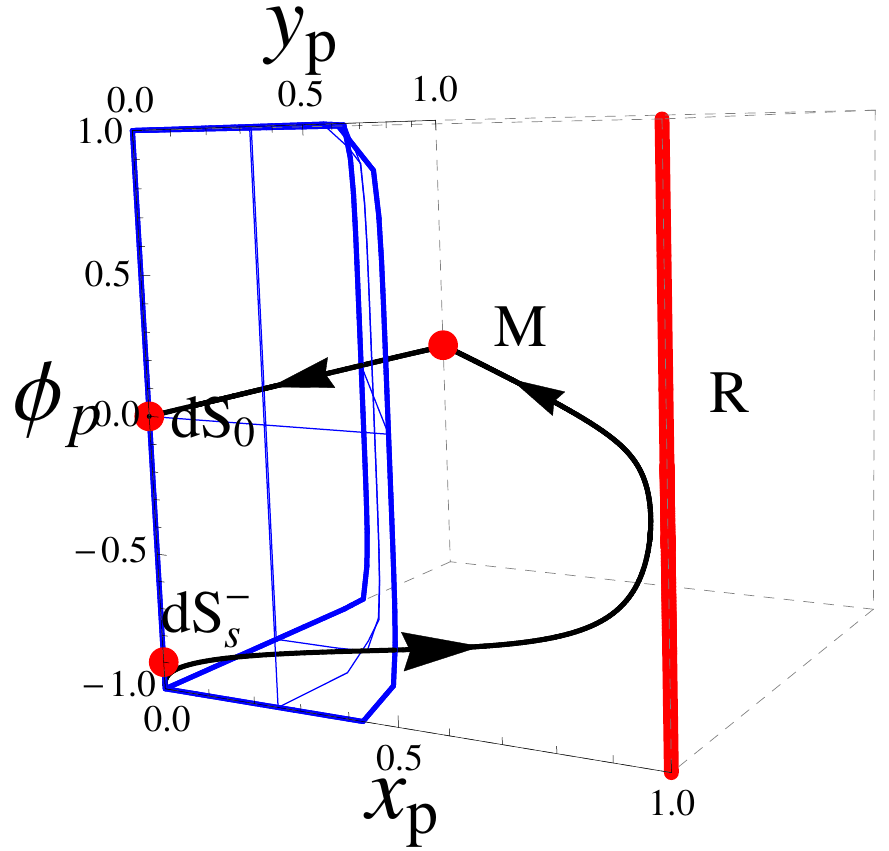}\label{poincare_3D_neg}}
	\qquad 
\subfigure[]{\includegraphics[width=5cm,height=5cm]{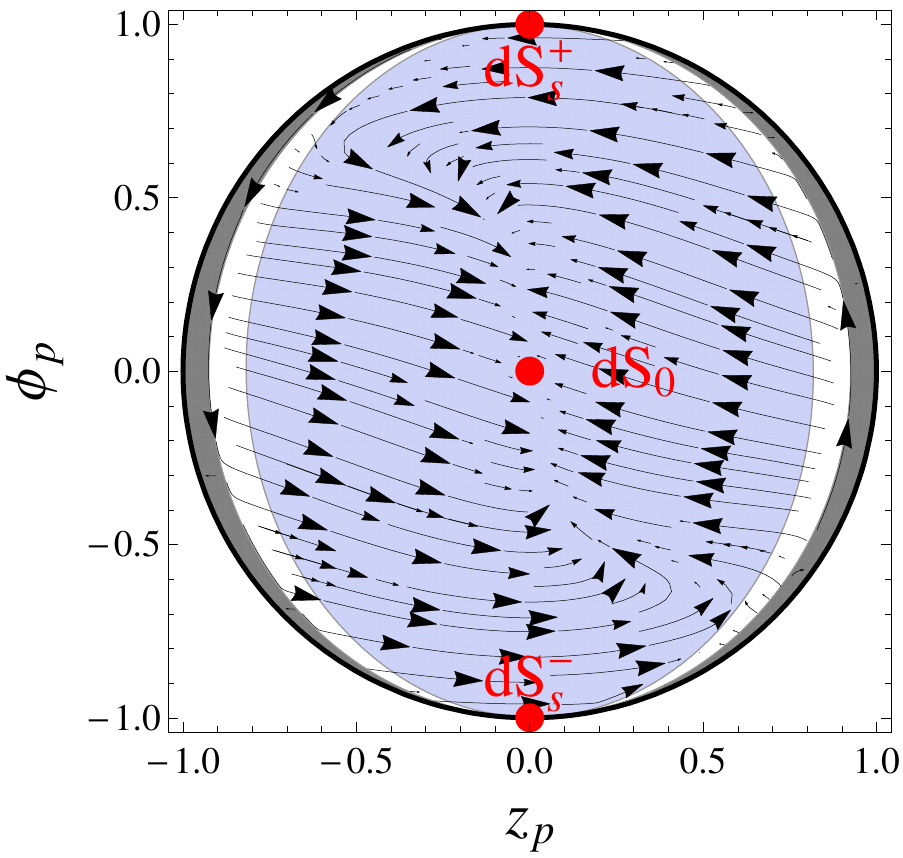}\label{poincare_2D_neg}}
	\\
\subfigure[]{\includegraphics[width=5cm,height=5cm]{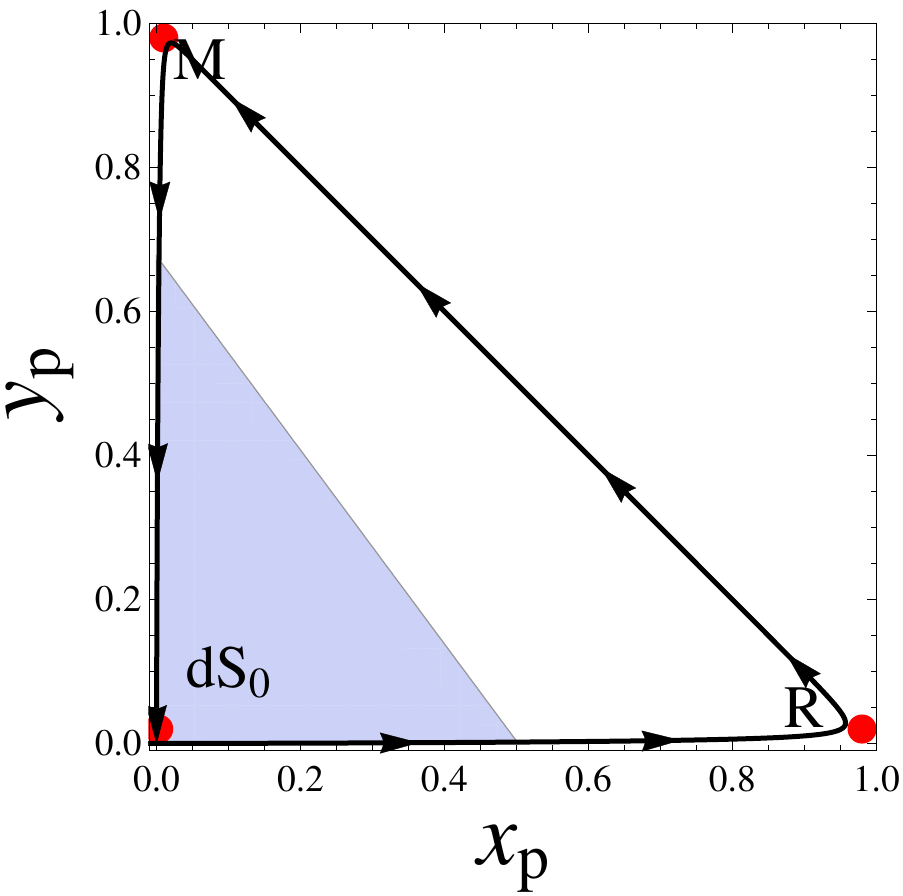}\label{acc_2d_neg}}
    \qquad
\subfigure[]{\includegraphics[width=5cm,height=5cm]{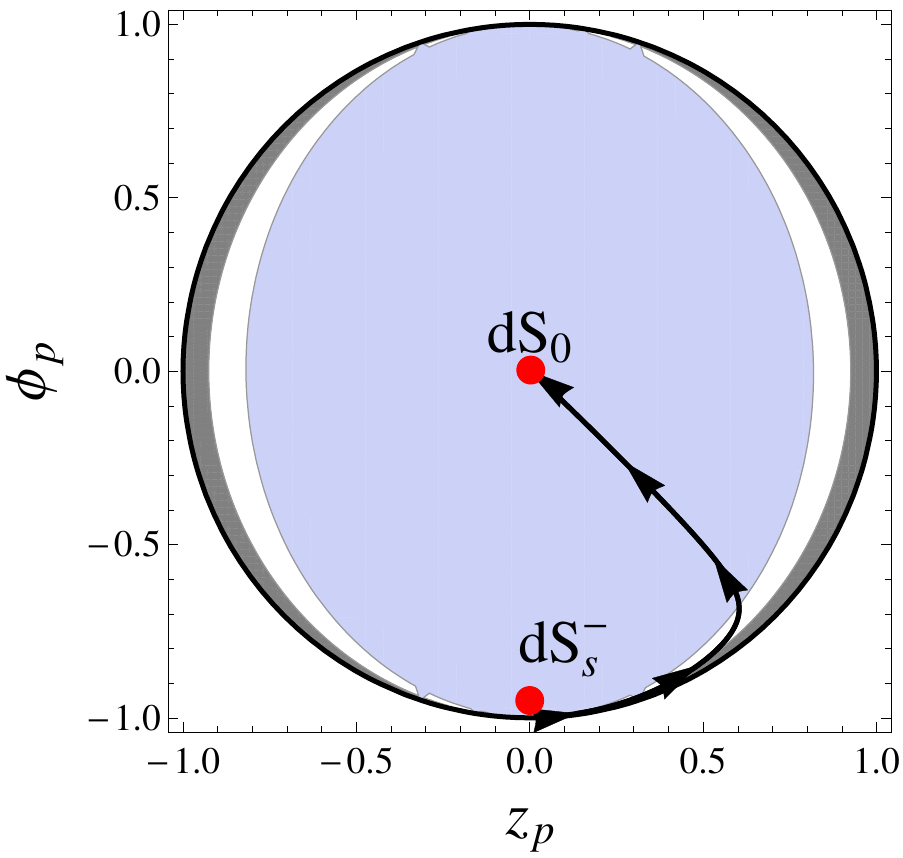}\label{2d_z_phi_pcre_neg}}
	\\
\subfigure[]{\includegraphics[width=5cm,height=4cm]{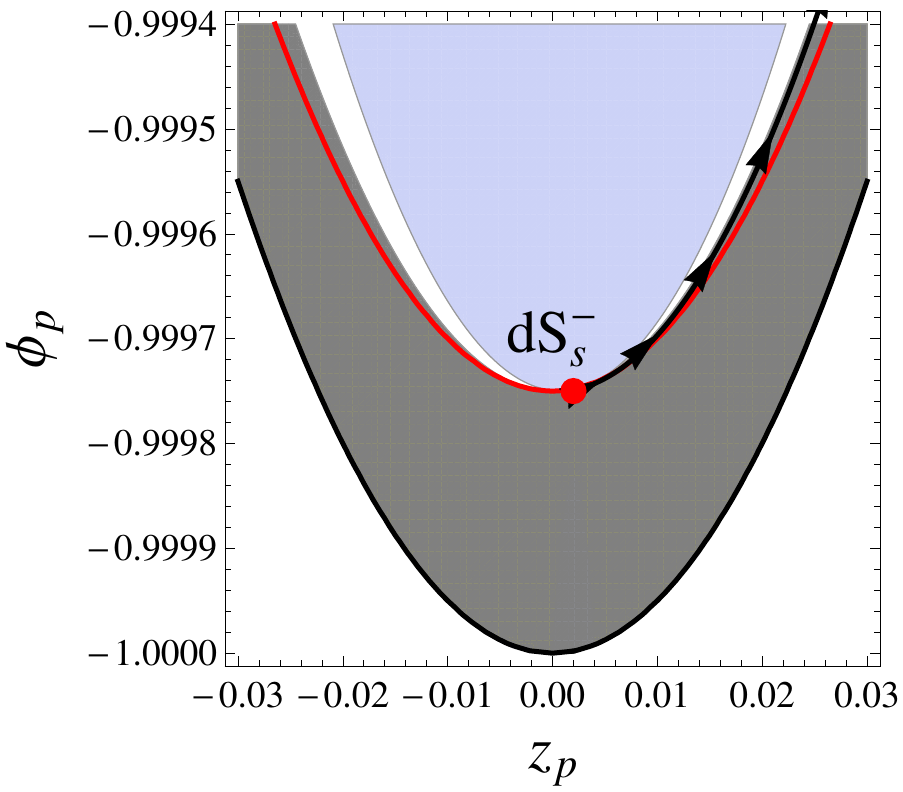}\label{2d_z_phi_pcre_neg_zoom}}
\caption{The dynamics in the system \eqref{dynsys x}-\eqref{dynsys phi} in the nonminimal quadratic coupling, quartic potential case \eqref{eq:nonminimal_quartic_potential} for an example of negative coupling ($\xi=-0.0005, V_0=0.001, \Lambda=0.005$). (a) An example phase trajectory in the $(x,y,\phi)$ dimensions, the region enclosed by a grid represents the accelerated expansion. (b) Evolution of  dust matter, radiation, and scalar field energy density parameters along with the effective equation of state along the depicted solution. (c) The same phase trajectory in the $(x_p,y_p,\phi_p)$ dimensions, where the range of the scalar field is compactified to $-1 \leq \phi_p \leq 1$. (d) Projection of the phase flow of the fully compactified system at the $x_p=y_p=0$ plane.
Here and on the following subplots the area shaded in blue represents the region of accelerated expansion, and the area in dark gray is ruled out by the Friedmann constraint \eqref{eq:Friedmann_constraint_nonmin_cpld}.
(e) Projection of the example trajectory to the $z_p=\phi_p=0$ plane, the shaded area represents the region of accelerated expansion. (f) Projection of the example trajectory to the $x_p=y_p=0$ plane. (g) Closer look at the projection of the example trajectory to the $x_p=y_p=0$ plane near the $dS_s$ fixed point, the red curve marks the values where effective gravitational constant would diverge.} 
\label{fig:xi_negative}
\end{figure}

For positive nonminimal coupling, $\xi>0$ the results are depicted on Fig.\ \ref{fig:xi_positive}. Here \ref{3D_x_y_phi_phi2_pow_pot} shows a finite patch of the $(x,y,\phi)$ subspace where a trajectory flows from the accelerated  region towards to the radiation domination point $R$, then turns to the matter domination point $M$, passes by the saddle de Sitter point $dS_0$, and finally ends up at de Sitter attractor $dS_+$. The evolution of the relative energy densities and effective barotropic index is given on Fig.\ \ref{parameters_phi2_pow_pot}. A hybrid plot \ref{poincare_3D} has the range of the scalar field compactified to a finite interval $-1\leq \phi_p \leq 1$, and we see how the trajectory actually goes back to the asymptotic fixed point $dS_\infty^-$. The boundary of accelerating region $w_{\rm eff}<-\tfrac{1}{3}$, indicated by the blue wireframe, of course includes all $dS$ points and extends further the $x-y$ directions near these points. This can explain the ``bump'' in the $w_{\rm eff}$  on Fig.\ \ref{parameters_phi2_pow_pot}. When the trajectory enters in the scalar field dominated era and first comes closer to $dS_0$ the effective barotropic index gets close to $-1$ as well, but while the trajectory then cruises away from $dS_0$ towards $dS_+$ the index bounces little bit back as the edge of the accelerating region is nearby for a while (see Fig. \ref{poincare_3D}). In fact, the final approach to $dS_+$ is marked by damped oscillations of $w_{\rm eff}$ below and above $-1$, i.e.\ periods of minute superacceleration. The first of such periods is barely noticeable on Fig.\ \ref{parameters_phi2_pow_pot}, while in each period to follow the superaccelerating effect is an order magnitude less pronounced than at the previous one. Such damped oscillations in the dynamics are a characteristic feature of the focus nature of the fixed point $dS_+$. On the plots \ref{poincare_2D} and \ref{2d_z_phi_pcre} of the $x_p=y_p=0$ slice of the fully compactified phase space \eqref{comp_phase}, the superaccelerating region $w_{\rm eff}<-1$ is shaded yellow. The plot \ref{2d_z_phi_pcre} shows a projection of the example trajectory onto this slice, and we see how it indeed lies on the superacceleration zone. The actual trajectory, however, does have nonvanishing radiation and matter, and it approaches $dS_+$ not exactly on the $x_p=y_p=0$ plane, but spirals down to it from these dimensions. At last, the Fig.\ \ref{acc_2d} shows a projection of the trajectory onto the $z_p=\phi_p=0$ slice of the compact phase space. We can compare with Fig.\ \ref{GR_phase_space_fig} and witness the correct sequence of cosmological eras to play out, viz.\ from a transient accelerating era, to radiation domination, to matter domination to late accelerating era of dark energy, as supplied by the fixed points $dS_\infty^- \rightarrow R \rightarrow M \rightarrow dS_+$. Due to the symmetry of the effective potential a similar cosmological sequence can be also realized by the fixed points $dS_\infty^+ \rightarrow R \rightarrow M \rightarrow dS_-$.

In the negative coupling case, $\xi<0$, the story is basically the same with few modifications, as depicted in Fig.\ \ref{fig:xi_negative}. There is just one regular de Sitter point $dS_0$ acting as the final attractor. Also, the range of the scalar field is constrained by $|\phi|<\frac{1}{\sqrt{-\xi}}$, and the role of the $dS$ saddle point is taken by $dS_s$, occurring at the boundary of the allowed region where the effective potential diverges. Indeed, on the plots, we can see how an example trajectory exhibits the eras of inflation, radiation domination, matter domination, and dark energy, as it flows guided by the fixed points $dS_s^- \rightarrow R \rightarrow M \rightarrow dS_0$. The final point in this particular example is a stable node and there is no region of superaccelerated expansion for this choice of parameters. (However, for large negative values of $\xi$ where $dS_0$ is a stable focus, there would also occur a superaccelerated region in the physical phase space and similar to the case where $\xi>0$, one can observe damped oscillatory behavior about $w_{\rm eff}=-1$.) Due to the symmetry of the effective potential, a similar cosmological sequence can be also realized by the fixed points $dS_s^+ \rightarrow R \rightarrow M \rightarrow dS_0$.

The last extra subplot \ref{2d_z_phi_pcre_neg_zoom} shows how the trajectory does indeed go back close to the $dS_s^-$ point and not at where would be $dS_\infty^-$. A red curve on that subfigure indicates the limit where the  effective gravitational constant diverges and the effective potential becomes singular. The unphysical region beyond that is not dynamically accessible from the physical part of the phase space. As explained before in the minimally coupled case, the saddle nature of $dS_s$ implies that only one particular phase trajectory originates from it, while all other trajectories in the vicinity converge to this particular solution. In principle, we could try to follow our example solution backwards beyond the immediate neighborhood of $dS_s^-$, but because of the singularity in the effective potential, the numerics becomes unreliable. In any case, the earlier adventures of the trajectory are not very relevant for subsequent history, as the $dS_s^-$ repulsive eigendirection guides all the nearby trajectories irrespective of their origin or way of approach. 

Although the two examples give the same qualitative picture of expansion history, the evolution of the effective gravitational constant \eqref{eq:G_eff} is radically different in them. In the positive coupling case ($\xi>0$), inflation takes place at very small values of $G_{\mathrm{eff}}$, which then grows around the radiation domination era and reaches a maximal value at the matter domination, but then drops again when the solution reaches $dS_+$. The drift of the effective gravitational constant in the matter domination era and later is rather restricted observationally, but can be suppressed if $\tfrac{\Lambda\xi^2}{V_0}\ll 1$, i.e.\ the fixed points $M$ and $dS_+$ are sufficiently closely aligned in the $\phi$ direction of the phase space. On the contrary, in the negative coupling case ($\xi<0$) inflation takes place at very large values of $G_{\mathrm{eff}}$, which then drops around the radiation domination era and reaches a minimal value at matter domination. Since the matter domination $M$ and final $dS_0$ occur at the same value of $\phi$, the evolution of the effective gravitational constant in the late universe would be marginally small for generic solutions (as they come close to the direct heteroclinic orbit from $M$ to $dS_0$).

Because unlike matter domination, radiation domination is not governed by a single fixed point but rather a line of fixed points extending in the $\phi$ direction, there is not much control on the value of the scalar field in this era, except that the evolution of $\phi$ in time should somewhat slow down at the peak of the radiation domination when a trajectory is most close to the line of $R$ fixed points. For an arbitrary trajectory, the scalar field value will evolve before and after the radiation domination peak. This has implications not only on the effective gravitational constant \eqref{eq:G_eff} but also on the relationship between the dust matter and radiation densities and the respective relative densities \eqref{eq:Omegas}, which also evolve with $\phi$. Therefore, it is interesting to speculate that the slightly varying scalar field value around the radiation domination era could be related to the observational puzzles of $H_0$ \cite{Aghanim:2018eyx, Wong:2019kwg, Reid:2019tiq, Verde:2019ivm} (for investigations in this direction see e.g.\ \cite{Ballesteros:2020sik, Braglia:2020iik, Ballardini:2020iws}) or even to the EDGES 21 cm line \cite{Bowman:2018yin}.

\section{Discussion}
\label{sec:discussion}

Contemporary cosmology is proliferated by a plethora of models proposed to describe the phenomena of inflation and dark energy. Most of these models are designed to have an effect at either high or low energy scales. However, a true model of fundamental physics must deliver correct cosmic history from early to late eras, and better exhibit this history as a generic property for a large class of solutions, not only certain particular ones. In the present work, we address this issue in the context of generic scalar fields minimally or nonminimally coupled to gravity, by employing the methods of dynamical systems with a useful combination of variables, whereby the principal cosmic eras appeared as fixed points with appropriate properties. 
To summarize, a viable cosmological model should have the following features:
\begin{enumerate}
\item inflation supported by a de Sitter fixed point with one repulsive eigendirection, i.e.\ either a regular $dS$ point at $\phi_*$ such that \eqref{dS fixed point condition} and \eqref{dS fixed point eigenvalue positive} are satisfied, or an asymptotic $dS_\infty$ or singular $dS_s$ described in Sec.\ \ref{sec:fixed_points}, these collect a large set of solutions and guide them to an inflationary path;
\item radiation domination fixed point which is a saddle, it is always guaranteed to exist;
\item dust matter domination fixed point which is also a saddle, for nonminimal coupling, it exists provided there is a value $\phi_m$ which satisfies \eqref{M fixed point condition};
\item final de Sitter fixed point manifesting as dark energy filled universe, when the condition \eqref{dS fixed point condition} is satisfied but the condition \eqref{dS fixed point eigenvalue positive} is not, all the  eigendirections are attractive as necessary.
\end{enumerate} 

It turns out that the main crux in confirming a good model is to check whether it has at least two de Sitter fixed points, one with a repulsive direction and another totally attractive. These fixed points can be understood to reside at the local or asymptotic maxima and minima of the effective potential \eqref{eq:V_eff}. If a model possesses the fixed points listed above, it has a good chance to reproduce qualitatively correct cosmic background evolution by a large set of its solutions. This list of fixed points provides the necessary barebones for a good model, but does not give an exhaustively sufficient condition, since there might be some other fixed points or features at the boundaries of the allowed phase space, which might alter or ruin the correct dynamics. However, as our variables allow to describe inflation in terms of a saddle de Sitter point, the question of how generic are the initial conditions for inflation and the eras that follow,  can now be mathematically framed using the tools of dynamical systems, left for further work.

It must be also noted that in our approach we considered all the cosmic components to be immutable and without interaction or change in character. This was a very strong simplifying assumption, which does not hold in real cosmology. First, the stage of inflation should end with the process of reheating where the scalar field kinetic energy gets transferred into radiation and matter fluids. Second, the massive matter particles which later behave as collisionless nonrelativistic dust were  at earlier times relativistic and behaved like radiation. These two phenomena should be included in the treatment of the full history of the universe as suitable interactions between the components, but we leave the engineering of these features for further work. When done properly the present set of fixed points should remain intact, but their basins of attraction would widen.

The aim of the current paper was to present the generic approach, while the systematic analysis of different  nonminimal coupling functions and potentials enabling viable cosmic history is left for another publication. In the models with nonminimal coupling, the effective gravitational constant becomes dynamical and brings in additional aspects to pay attention to. Of course, background history is just one facet of observational viability, the other important issue is the treatment of perturbations, which remains beyond the scope of the current work. Besides cosmology, a correct model of gravity must also pass the tests coming from the Solar System, black holes, neutron stars, gravitational waves, etc. Still, for any scalar-tensor model of interest the results presented here do offer a  quick viability check of basic cosmology, and allow to focus on next issues.

Finally, let us emphasize that the  method presented here for scalar-tensor type of theories is completely general and can be easily adjusted to analyze other classes of theories as well. It should be rather straightforward to treat models with a scalar field nonminimally coupled to Palatini Ricci scalar \cite{Wang:2013wuf}, Gauss-Bonnet invariant \cite{Nojiri:2005vv}, torsion scalar \cite{Geng:2011aj,Hohmann:2018rwf}, nonmetricity scalar \cite{Jarv:2018bgs}, etc., in a rather analogous manner, while in principle any more complex generalizations involving derivative couplings, higher derivative terms, multiple fields, and so on, can be subject to similar analysis as well.

\begin{acknowledgments}
The authors are grateful to Alexey Toporensky for insightful discussions. During the research LJ was funded by the Estonian Research Council through the projects IUT02-27, PUT790, and PRG356, as well as by the European Regional Development Fund through the Center of Excellence TK133 “The Dark Side of the Universe”. JD was supported by the Core Research Grant of SERB, Department of Science and Technology India (File No. CRG $\slash 2018 \slash 001035$) and the Associate program of IUCAA.
\end{acknowledgments}

\appendix

\section{Stability analysis for non-hyperbolic point dS}
\label{CMTdS1}
In this appendix, we apply the method of the center manifold theory  in order to study the stability of the non-hyperbolic point dS $(0,0,0,\phi_*)$ for $f(\phi)=0$ and potential $V(\phi)$ such that $V_{,\phi}(\phi_*)=0$ and $V_{,\phi\phi}(\phi_*)=0$. Under this method, we introduce a new set of variables $(X,Y,Z,\Phi)$, expressed in terms of the original set
of variables $(x,y,z,\phi)$ as
\[\left(\begin{array}{c}
X\\
Y\\
Z\\
\Phi \end{array} \right)=\left(\begin{array}{cccc}
1   & 0 & 0&0 \\
0    &1   & 0&0\\
0    & 0 & -\frac{1}{3}&0 \\
0    & 0 & \frac{1}{3}&1 \\ \end{array} \right) \left(\begin{array}{c}
x\\
y\\
z\\
\phi \end{array} \right) \,. \]
 The system of equations \eqref{dynsys x}-\eqref{dynsys phi} can be recast in terms of these new set of variables as
\[\left(\begin{array}{c}
X'\\
Y'\\
Z'\\
\Phi' \end{array} \right)=\left(\begin{array}{cccc}
-4  & 0 & 0&0  \\
0  & -3  & 0 &0 \\
0  & 0 & -3& 0 \\
0  & 0 & 0 & 0   \end{array} \right) \left(\begin{array}{c}
X\\
Y\\
Z\\
\Phi \end{array} \right)+\left(\begin{array}{c}
g_1\\
g_2\\
g_3\\
f \end{array} \right),\]
where $g_1,\,g_2,\,g_3,\, f$ are polynomials of degree greater than 2 in $(X,\,Y,\,Z,\Phi)$, with
\begin{eqnarray}
g_1&=&9\,{Z}^{2}X+4\,{X}^{2}+3\,XY\\
g_2&=&9\,{Z}^{2}Y+4\,XY+3\,{Y}^{2}\\
g_3&=&\frac{9}{2}\,{Z}^{3}-\frac{3}{2}\,{\frac {\mbox {D} V\left( Z+\Phi+
\phi_* \right) {Z}^{2}}{V\left( Z+\Phi+\phi_* \right) }}+2\,XZ+\frac{3}{2}\,Z
Y \nonumber \\
&&  -{\frac {\mbox {D} V  \left( Z+\Phi+\phi_* \right) X}{
V \left( Z+\Phi+\phi_* \right) }}-{\frac {\mbox {D} V 
 \left( Z+\Phi+\phi_* \right) Y}{V \left( Z+\Phi+\phi_* \right) }}+{
\frac {\mbox {D} V \left( Z+\Phi+\phi_* \right) }{V
 \left( Z+\Phi+\phi_* \right) }}
\\
f&=&-g_3
\end{eqnarray}
where $V$ is a function of new variable $Z+\Phi$ (which is $\phi$ in terms of old variables) and D denotes derivative of $V$ with respect to $Z+\Phi$. The local center manifold is given by
\begin{equation}
\left\lbrace (X,Y,Z) : X=h_1(\Phi), Y=h_2(\Phi), Z=h_3(\Phi), h_i(0)=0,  Dh_i(0)=0 , i=1,2,3 \right\rbrace,
\end{equation}
where $h_1$, $h_2$ are approximated as
\begin{align}
h_1(\Phi)=a_2 \Phi^2+a_3 \Phi^3+\mathcal{O}(\Phi^4),\\
h_2(\Phi)=b_2 \Phi^2+b_3 \Phi^3+\mathcal{O}(\Phi^4),\\
h_2(\Phi)=c_2 \Phi^2+c_3 \Phi^3+\mathcal{O}(\Phi^4),
\end{align}
respectively. The invariant property of the center manifold implies that the function $\mathbf{h}$ has to satisfy a quasilinear partial differential equation given by
\begin{equation}\label{quasi_C}
D \mathbf{h(\Phi)}\left[A \Phi+\mathbf{F}(\Phi,\mathbf{h}(\Phi))\right]-B
\mathbf{h}(\Phi)-\mathbf{g}(\Phi,\mathbf{h}(
\Phi))=\mathbf{0} \,,
\end{equation}
with
\[\mathbf{h}=\left(\begin{array}{c}
h_1\\
h_2\\
h_3\ \end{array} \right),~~ \mathbf{g}=\left(\begin{array}{c}
g_1\\
g_2\\
g_3 \end{array} \right),~~~~~ \mathbf{F}=f, ~~~~~B= \left(\begin{array}{ccc}
 -4 & 0&0 \\
 0 &-3&0\\
 0&0&-3 \end{array} \right),~~~~~ A=0. \]
By comparing the coefficients of all powers of $\Phi$ in Eqn. \eqref{quasi_C}, we obtain the constants $a_2$,
$a_3$, $b_2$, $b_3$, $c_2$, $c_3$ as
\begin{gather}
    a_2=0 \,,\quad a_3=0\,,\qquad b_2=0\,, \qquad b_3=0\,,\qquad c_{2}=\frac{1}{6}\,\Big[\frac {  D^{ \left( 3 \right) } V }{V}\Big]_{\Phi=\phi_*}
\,,\qquad c_3=\frac{1}{18}\Big[ \frac {2\, \left( D^{ \left( 3 \right) } V\right)  ^{2}+V \left( D^{ \left( 4
 \right) }  V \right)}{ V ^{2}}\Big]_{\Phi=\phi_*}.
\end{gather}
The dynamics of the reduced system is eventually determined by the equation
\begin{equation}
\Phi'=A\,\Phi+\mathbf{F}(\Phi,\mathbf{h}(\Phi)),
\end{equation}
and hence
\begin{align}
\Phi'=-\frac{1}{2}\Big[\frac {  D^{ \left( 3 \right) } V }{V}\Big]_{\Phi=\phi_*}{\Phi}^{2}-\frac{1}{6}\Big[ \frac {\left( D^{ \left( 3 \right) } V\right)  ^{2}+V \left(D^{ \left( 4
 \right) }  V \right) }{ V ^{2}}\Big]_{\Phi=\phi_*}{\Phi}^{3}+\mathcal{O}(\Phi^4).
\end{align}
Therefore,  if $\Big[\frac {  D^{ \left( 3 \right) } V }{V}\Big]_{\Phi=\phi_*}=0$, this point is stable if $\Big[\frac {  D^{ \left( 4 \right) } V }{V}\Big]_{\Phi=\phi_*}>0$ and saddle if $\Big[\frac {  D^{ \left( 4 \right) } V }{V}\Big]_{\Phi=\phi_*}<0$. It is also saddle if $\Big[\frac {  D^{ \left( 3 \right) } V }{V}\Big]_{\Phi=\phi_*} \neq 0$.


%

\end{document}